\documentclass[showpacs,preprintnumbers,amsmath,amssymb]{revtex4}

\usepackage{graphicx}
\usepackage{dcolumn}
\usepackage{bm}
\usepackage{xcolor}
\usepackage{placeins}

\newcommand{\revblue}[1]{\textcolor{blue}{#1}}

\newcommand{\bq}{\begin{equation}}
\newcommand{\eq}{\end{equation}}
\newcommand{\bqa}{\begin{eqnarray}}
\newcommand{\eqa}{\end{eqnarray}}
\newcommand{\nn}{\nonumber \\}

\def\be{\begin{equation}}
\def\ee{\end{equation}}
\def\bea{\begin{eqnarray}}
\def\eea{\end{eqnarray}}
\def\bnn{\begin{eqnarray*}}
\def\enn{\end{eqnarray*}}

\begin{document}

\title{Functional Renormalization Group as a Ricci Flow: An \(\mathcal{F}\)-Entropy Perspective on Information Metric Dynamics}

\author{Jin Mo Bok$^{1}$ and Ki-Seok Kim$^{1,2}$}

\affiliation{$^{1}$Department of Physics, POSTECH, Pohang, Gyeongbuk 37673, Korea}
\affiliation{$^{2}$Asia Pacific Center for Theoretical Physics (APCTP), Pohang, Gyeongbuk 37673, Korea}

\email[Jin Mo Bok: ]{jinmobok@postech.ac.kr}
\email[Ki-Seok Kim: ]{tkfkd@postech.ac.kr}

\date{\today}

\begin{abstract}
We establish a geometric correspondence between the Functional Renormalization Group (FRG) and a Ricci flow modified by a potential-driven diffeomorphism.
By rewriting the Polchinski exact RG equation as an infinite-dimensional Fokker--Planck equation for field-distribution functionals, we show how a probability flow driven by a ``thermodynamic'' free-energy functional induces the evolution of the Fisher information metric on the coupling-constant manifold.
Using the continuous scale-dissipation rate of this free-energy functional, we construct an RG-flow entropy functional that provides an infinite-dimensional counterpart of Perelman's \(\mathcal{F}\)-entropy.
The parametric Hessian of this RG-flow entropy then encodes the scale deformation of the Fisher information metric, thereby linking the JKO--Wasserstein flow in field-configuration space to the geometry of the coupling-constant manifold.
An emergent scalar information potential \(\Phi\) encodes the potential-driven diffeomorphism component, restoring the tensorial form of the flow under reparametrizations of the coupling coordinates.
In this representation, the successive integration of high-energy degrees of freedom effectively smooths out the curvature of the information manifold, so that RG fixed points are realized as steady Ricci soliton equilibria.
These results connect quantum field theory, optimal transport, and Perelman's theory of geometric evolution, providing a geometric framework for characterizing the stability, universality, and topological structure of quantum field theories.
\\
\end{abstract}


\maketitle

\section{Introduction}

The Renormalization Group (RG), introduced by Wilson and Wegner \cite{Wilson1971, Wegner1972}, provides a systematic framework for describing how effective quantum theories change under coarse graining and how macroscopic phenomena emerge from microscopic degrees of freedom.
Within this broad framework, the Functional Renormalization Group (FRG) provides nonperturbative exact flow equations, most prominently the Wetterich equation for the effective average action \(\Gamma_\Lambda\) \cite{Wetterich1993}.
However, a general information-geometric principle that connects exact RG trajectories to the metric and curvature dynamics of the coupling-constant manifold has not yet been fully established.

A natural language for such a principle is information geometry, which endows a parameterized family of probability distributions with a Riemannian structure through the Fisher information metric, thereby quantifying the statistical distinguishability of neighboring effective theories \cite{Rao1945,Amari2000,Amari2016,Ay2017,Nielsen2022}.
Related information-geometric and thermodynamic approaches have been used to interpret RG trajectories in theory space as geodesic motion, entropy-producing flows, or gradient flows associated with monotone functionals \cite{Zamolodchikov1986,Cardy1988,Shore1991,Kar2001,Ruppeiner1979,Gaite1996,Crooks2007,Kluth2024,RG_Monotonicity_NEQ}.
Coarse graining has also been interpreted in terms of information loss and statistical inference, and RG transformations have been formulated as quantum channels whose contraction of distinguishability induces an information-geometric structure \cite{Apenko2012,BenyOsborne2015PRA,BenyOsborne2015NJP}. More recently, an exact FRG evolution equation has been derived for the divergence functional, while the information geometry of Euclidean quantum fields has been developed at the functional level in terms of metrics, connections, and covariant derivatives \cite{Floerchinger2023,Floerchinger2024}.

On the geometric-analysis side, Hamilton's Ricci flow and Perelman's \(\mathcal{F}\)-entropy provide an entropy-gradient formulation of curvature evolution, in which the \(\mathcal{F}\)-functional is monotone along the modified Ricci-flow dynamics \cite{Hamilton1982, Perelman2002, Chow2004, Kleiner2008,Morgan2007,Topping2006}.
A well-known physical precedent for relating RG flow to curvature evolution arises in nonlinear sigma models, where the one-loop beta function of the target-space metric is proportional to the target-space Ricci tensor \cite{Friedan1980, Callan1985, Fradkin1985, Tseytlin2007, Carfora2010, Codello2009,RG_Flow_Nonperturbative_String}.
However, this target-space relation does not by itself provide a curvature-flow formulation for the Fisher information metric on the coupling-constant manifold.
Recent optimal-transport approaches have shown that Polchinski's exact RG equation can be reformulated as a Wasserstein gradient flow of a field-theoretic relative entropy \cite{CotlerRezchikov2023}.
What remains to be developed is a tensorial geometric description of the induced Fisher-metric flow, including the intrinsic curvature of this information manifold and the diffeomorphism component associated with reparametrizations of the coupling coordinates.

In this paper, we bridge these approaches by starting from the Polchinski form of the exact RG and constructing a modified Ricci-flow representation of the induced Fisher-metric dynamics.
Specifically, we formulate this exact RG flow as a functional JKO--Wasserstein gradient flow of probability functionals over field configurations, driven by a ``thermodynamic'' free-energy functional \cite{CotlerRezchikov2023}.
From this construction, we propose a modified Ricci-flow equation for the Fisher information metric on the coupling-constant manifold: its scale derivative decomposes into the intrinsic Ricci curvature and the covariant Hessian of an emergent scalar information potential \(\Phi\).
Although \(\Phi\) is determined implicitly rather than solved in closed form, its covariant Hessian generates the potential-driven diffeomorphism term that absorbs the coordinate-dependent components of the Fisher-metric flow and restores covariance under reparametrizations of the coupling coordinates.

We further construct an RG-flow entropy functional from the continuous scale-dissipation rate of the same ``thermodynamic'' free-energy functional, showing that the renormalization-induced deformation of the Fisher metric is governed by the parametric Hessian of this entropic landscape.
Consequently, these results place the nonperturbative exact RG flow within Perelman's theory of geometric evolution, providing a geometric framework for characterizing RG fixed points as steady Ricci soliton equilibria of the information manifold.

\section{From Fokker-Planck equation to Ricci flow}
\label{FPE_RF}

\subsection{From Fokker-Planck equation to information metric flow}

To describe the dynamical evolution of a statistical ensemble, we consider the Fokker-Planck equation \cite{Risken1996, Gardiner2009, Zwanzig2001, ZinnJustin2021}, which governs the spatio-temporal evolution of the probability density function \(p(x, t; \theta^a)\):
\bqa
&& \partial_{t} p(x,t;\theta^{a})
= - \partial_{i} \Big( A^{i}(x;\theta^{a}) p(x,t;\theta^{a}) \Big)
+ \partial_{i} \Big( B^{ij}(x;\theta^{a}) \partial_{j} p(x,t;\theta^{a}) \Big) .
\label{FP_Eq}
\eqa
Here, the drift vector \(A^{i}\) represents the deterministic macroscopic flow of the system, and the diffusion tensor \(B^{ij}\) accounts for stochastic fluctuations. This Fokker-Planck framework is introduced based on the structural correspondence between statistical diffusion and the RG flow, where the successive integration of high-energy modes operates as a coarse-graining process. Consequently, Eq. (\ref{FP_Eq}) determines the scale-dependent evolution of the information metric \(g_{ab}\), providing the initial framework for the geometric mapping to the Ricci flow.

To analyze the geometric structure of the parameter space associated with \(p(x, t; \theta^a)\), we employ the Fisher information metric \(g_{ab}\), defined as \cite{Rao1945, Amari2000, Jeffreys1946}:
\bqa
&& g_{ab}(t,\theta^{a})
= \int d^{d} x\, p(x,t;\theta^{a})
\partial_{a} \ln p(x,t;\theta^{a})
\partial_{b} \ln p(x,t;\theta^{a}) .
\label{Fisher_IM}
\eqa
This metric quantifies the distinguishability of probability distributions under an infinitesimal variation of the parameters \(\theta ^{a}\), endowing the parameter manifold with a Riemannian structure that reflects the fluctuations of the system.

When the probability density \(p(x, t; \theta^a)\) evolves according to Eq. (\ref{FP_Eq}), the time evolution rate of the Fisher information metric, \(\dot{g}_{ab}\), is derived from the continuity structure of the probability current. Following geometric formulations of stochastic systems \cite{Amari1985, ItoDechant2020}, the metric deformation equation is given by:
\bqa
&& \dot{g}_{ab}(t,\theta^{a})
= \int d^{d} x\, p(x,t;\theta^{a})
\Big[
\Big(\partial_{a} v^{i}(x,t;\theta^{a}) \Big)
\partial_{i} \Big( \partial_{b} \ln p(x,t;\theta^{a}) \Big)
+
\Big(\partial_{b} v^{i}(x,t;\theta^{a}) \Big)
\partial_{i} \Big( \partial_{a} \ln p(x,t;\theta^{a}) \Big)
\Big] .
\label{Metric_Eq_1st}
\eqa
where $v^i(x, t; \theta^a)$ denotes the effective velocity field of the probability current, defined as:
\bqa
&& v^{i}(x,t;\theta^{a})
= A^{i}(x;\theta^{a})
- \frac{1}{p(x,t;\theta^{a})}
B^{ij}(x;\theta^{a}) \partial_{j} p(x,t;\theta^{a}) .
\label{Eff_Force}
\eqa
Equation (\ref{Eff_Force}) defines the net probability current velocity by balancing the drift \(A^{i}\) against the diffusion-induced flow. Accordingly, Eq. (\ref{Metric_Eq_1st}) shows that the evolution of the information metric is determined by the gradients of this velocity field coupled with the informational score functions. This relation indicates that local information dissipation during drift and diffusion processes results in a global geometric deformation of the parameter space, establishing the mathematical basis for a comparison with the canonical Ricci flow. For this formulation, we assume that the probability density \(p(x, t; \theta^a)\) belongs to a smooth, rapidly decreasing continuous class, where parameter derivatives decay faster than any polynomial at the boundary, ensuring that all boundary terms from integration by parts vanish.

\subsection{From the information metric flow to the Ricci flow}

Ricci flow evolves a Riemannian metric according to \(\partial_t g_{ab}=-2R_{ab}\), where \(R_{ab}\) is the Ricci tensor constructed from \(g_{ab}\). It describes how the local distance structure of a manifold changes in response to its intrinsic curvature. For the Fisher metric, these distances quantify the statistical distinguishability of neighboring probability distributions on the parameter manifold. Introductory accounts of the geometric ideas behind Ricci flow can be found in Refs.~\cite{Chow2004,Topping2006}.

Although Eq.~\eqref{Metric_Eq_1st} describes the Fisher-metric deformation induced by Fokker--Planck dynamics, this statistically generated flow cannot be identified directly with the canonical Ricci flow without separating the coordinate-dependent part of the deformation.
The canonical Ricci flow is governed solely by the intrinsic Riemannian curvature of the manifold, whereas the Fisher-metric flow derived from the statistical dynamics generally contains coordinate-dependent deformation terms tied to the choice of coordinates on the parameter manifold.
These coordinate-dependent terms reflect the freedom to reparametrize the statistical manifold. Since the Fisher metric measures distinguishability among probability distributions rather than the particular coordinate labels used to describe them, reparametrizations of the coupling coordinates appear as Lie-derivative contributions to the metric flow rather than as intrinsic curvature evolution.
Consequently, Eq.~\eqref{Metric_Eq_1st} should not be identified directly with the canonical Ricci flow. Before such a comparison can be made, the coordinate-dependent Lie-derivative contribution associated with reparametrizations of the parameter manifold must be separated from the intrinsic curvature contribution. This leads naturally to a modified Ricci-flow description with a diffeomorphism term \cite{Carfora2010, Tseytlin1987, Headrick2006}.
Appendix A gives the explicit comparison between the Fisher-metric flow and the Ricci tensor constructed from \(g_{ab}\), and derives the diffeomorphism constraint required for the modified Ricci-flow form.

This structure is analogous to gauge fixing in gauge field theories and to the DeTurck modification of Ricci flow, where an auxiliary diffeomorphism term is introduced to control coordinate freedom and, in the Ricci-flow case, to obtain a strictly parabolic evolution equation \cite{Headrick2006, DeTurck1983, Chow2007}.
To make this identification geometrically consistent, the information-metric dynamics are embedded in a modified Ricci-flow framework that includes an explicit diffeomorphism term.
On the parameter manifold, an infinitesimal reparametrization generated by a vector field \(X^{a}\) changes the metric tensor by its Lie derivative, \( \mathcal{L}_X g_{ab} = \nabla_a X_b+\nabla_b X_a . \)
When this vector field is generated by a scalar information potential \(\Phi\) according to the convention \(X_a=-\nabla_a\Phi\), the Lie-derivative term becomes minus twice the covariant Hessian of \(\Phi\), \(\mathcal{L}_X g_{ab} =-2\nabla_a\nabla_b\Phi\).
Because the coordinate-dependent components of the Fokker--Planck-induced Fisher-metric flow have this Lie-derivative structure, they are absorbed into the potential-driven diffeomorphism generated by \(\Phi\).
Accordingly, the information-metric evolution takes the modified Ricci-flow form
\bqa
&& \dot{g}_{ab}(t,\theta)
= -2R_{ab}(t,\theta)-2\nabla_a\nabla_b\Phi(t,\theta) .
\label{Modified_Ricci_Flow_1st}
\eqa
Here \(R_{ab}\) is the Ricci tensor constructed from the Fisher information metric \(g_{ab}\), while \(\nabla_a\nabla_b\Phi\) is the covariant Hessian of the scalar information potential.

The information potential \(\Phi\) is not an arbitrary auxiliary field.
At the scalar trace level, it is fixed by the modified Ricci-flow relation
together with the Fisher-metric deformation. With
\(\Box_g\equiv g^{ab}\nabla_a\nabla_b\) and
\(R\equiv g^{ab}R_{ab}\), one obtains
\bqa
&& R(t,\theta)+\Box_g\Phi(t,\theta)
=
-g^{ab}(t,\theta)
\int d^d x\,p(x,t;\theta)
\bigl(\partial_a v^i(x,t;\theta)\bigr)
\partial_i\bigl(\partial_b\ln p(x,t;\theta)\bigr).
\label{Phi_Trace_Equation}
\eqa
Equation~\eqref{Phi_Trace_Equation} is obtained by contracting the modified Ricci-flow equation with \(g^{ab}\) and using the symmetry \(g^{ab}=g^{ba}\) in Eq.~\eqref{Metric_Eq_1st}.
For a nondegenerate Fisher metric, \(\Box_g\) is an elliptic Laplace--Beltrami operator, so Eq.~\eqref{Phi_Trace_Equation} is a standard Poisson-type equation for \(\Phi\), with the usual existence and uniqueness conditions determined by the chosen boundary or normalization data~\cite{Evans2010}.

\subsection{$\mathcal{F}-$entropy functional}

To describe the statistical state and its geometric evolution, we introduce the ``thermodynamic'' free energy functional $\mathcal{E}[p]$, which governs the optimal transport dynamics, and the $\mathcal{F}$-entropy functional acting as the information-geometric Lyapunov potential analogous to Perelman's energy functional~\cite{Perelman2002}:
\begin{equation}
\mathcal{E}[p(x,t;\theta^{a})]
= \int d^{d} x\, p(x,t;\theta^{a})
\Big( \ln p(x,t;\theta^{a}) + V(x;\theta^{a}) \Big) ,
\label{Correct_Free_Energy}
\end{equation}
where $V(x;\theta^a)$ defines the potential landscape. The drift term $A_{i}$ and the diffusion tensor $B^{ij}(x;\theta^a)$ of the Fokker-Planck equation determine the trajectory along this landscape via the relation:
\begin{equation}
A^{i}(x;\theta^{a})
= - B^{ij}(x;\theta^a)\partial_{j} V(x;\theta^{a}) .
\label{Drift_Potential_Relation}
\end{equation}
Based on these definitions, the Fokker-Planck equation can be reformulated into a Wasserstein gradient flow representation through the functional variation of the free energy $\mathcal{E}[p]$, in accordance with the JKO framework for dissipative systems~\cite{Jordan1998, Otto2001, Villani2009}:
\begin{equation}
\partial_{t} p(x,t;\theta^{a})
=
\partial_{i}
\Big\{
p(x,t;\theta^{a}) B^{ij}(x;\theta^{a})
\partial_{j}
\Big(
\frac{\delta \mathcal{E}[p(x,t;\theta^{a})]}
{\delta p(x,t;\theta^{a})}
\Big)
\Big\} .
\label{FPE_Gradient_Flow_Corrected}
\end{equation}
Equation~(\ref{FPE_Gradient_Flow_Corrected}) shows that the temporal evolution of the probability distribution $p$ follows the gradient of the free energy landscape.

The rate of this energy dissipation defines the $\mathcal{F}$-entropy functional:
\begin{equation}
\mathcal{F}[p(x,t;\theta^{a})]
\equiv -\frac{1}{2}\frac{d\mathcal{E}}{dt}
=
\int d^{d} x\, p(x,t;\theta^{a})
\Big(
\frac{1}{2} B^{ij}(x;\theta^{a})
\partial_{i} \ln p(x,t;\theta^{a})
\partial_{j} \ln p(x,t;\theta^{a})
+ \tilde{V}(x;\theta^{a})
\Big) .
\label{W_Entropy_FKE_Corrected}
\end{equation}
where the effective potential $\widetilde{V}(x; \theta^a)$ is given by:
\begin{equation}
\widetilde V(x;\theta^a)
=
\frac12
B^{ij}(x;\theta^a)
\partial_i V(x;\theta^a)
\partial_j V(x;\theta^a)
+
\partial_i A^i(x;\theta^a).
\label{Effective_Potential_FP}
\end{equation}

By applying the dual mapping between optimal transport dynamics and information-geometric variations on the parameter space~\cite{Amari1985,Lott2009,Crooks2007}, the Fisher-metric flow induced by Eq.~\eqref{Metric_Eq_1st} can be related to the second parameter variation of the statistical \(\mathcal{F}\)-entropy. Keeping the ``off-shell'' residual sector explicit, this relation takes the form:
\begin{equation}
\dot{g}_{ab}(t,\theta^{a})
=
-2
\frac{\partial^{2}\mathcal{F}[p(t,x;\theta^{a})]}
{\partial\theta^{a}\partial\theta^{b}}
+
2\int d^d x\,
\left[
(\partial_a\partial_b p)\mathcal{H}
+
p\mathcal{R}_{ab}
\right].
\label{Ricci_Flow_Gradient_Flow_1st_Corrected}
\end{equation}
Equation~\eqref{Ricci_Flow_Gradient_Flow_1st_Corrected} shows that the Fisher-metric deformation is governed by the second-order parametric derivative of the $\mathcal{F}$-entropy functional.
Here \(\mathcal{H}\) is the potential-constraint block, and \(p\mathcal{R}_{ab}\) is the weighted parametric remnant generated by the direct second parameter differentiation of \(\mathcal{F}\).
The explicit forms of \(\mathcal{H}\) and \(p\mathcal{R}_{ab}\), together with their cancellation mechanism on the stationary zero-current profile, are given in Appendix~\ref{app:residual-cancellation}.
In that sector, \(\mathcal{H}\) vanishes pointwise and the weighted remnant \(p\mathcal{R}_{ab}\) reduces to a total divergence; under the same boundary conditions used in the integrations by parts, the integrated residual term vanishes and Eq.~\eqref{Ricci_Flow_Gradient_Flow_1st_Corrected} reduces to the Hessian matching relation.

To establish the geometric relation between the metric evolution and the Ricci flow, we derive the condition relating the curvature of the parameter space to the Hessian of the entropy functional.
Combining the modified Ricci flow equation with the entropic identity in Eq.~(\ref{Ricci_Flow_Gradient_Flow_1st_Corrected}) under the potential ansatz $X_{a} = -\nabla_{a} \Phi$ reorganizes the connection components into a generalized soliton equation.
Imposing a torsion-free Levi-Civita connection yields the geometric identity:
\begin{equation}
R_{ab}(t,\theta^{a})
+
\nabla_{a}\nabla_{b}\Phi(t,\theta^{a})
=
\frac{\partial^{2}\mathcal{F}[p(t,x;\theta^{a})]}
{\partial\theta^{a}\partial\theta^{b}}
-
\int d^d x\,
\left[
(\partial_a\partial_b p)\mathcal{H}
+
p\mathcal{R}_{ab}
\right].
\label{To_Determine_Phi_Corrected}
\end{equation}
Equation~\eqref{To_Determine_Phi_Corrected} is the curvature relation before restricting to the compatible Fisher--Wasserstein sector.
The Wasserstein interpretation clarifies the geometric meaning of this compatible sector.
In the JKO--Otto formulation, tangent vectors to the probability space are represented by gradient transport currents.
A general off-shell probability current may contain an additional non-gradient, or rotational, component, which measures the departure from the intrinsic Wasserstein-gradient sector.
The compatible sector considered here restricts the probability transport to this gradient sector; hence the non-gradient contribution is absent, while the residual terms in Eqs.~\eqref{Ricci_Flow_Gradient_Flow_1st_Corrected} and \eqref{To_Determine_Phi_Corrected} are removed by the stationary zero-current cancellation described in Appendix~\ref{app:residual-cancellation}.
This formulation corresponds to the geometric embedding of Fokker-Planck dynamics within a Ricci flow background~\cite{Carfora2007, Lott2009, Villani2009}, identifying the optimal transport relaxation of the ensemble with Perelman's steady Ricci soliton evolution.

Based on the relation in Eq.~(\ref{To_Determine_Phi_Corrected}), the global geometric \(\mathcal{F}\)-entropy functional on the parameter space is defined as:
\begin{equation}
\mathcal{F}[g_{ab}(t,\theta^{a})]
=
\int d^{N} \theta \sqrt{g(t,\theta^{a})} e^{- \Phi(t,\theta^{a})}
\Big\{
R(t,\theta^{a})
+
g^{ab}(t,\theta^{a})
\partial_{a} \Phi(t,\theta^{a})
\partial_{b} \Phi(t,\theta^{a})
\Big\} .
\label{Perelman_F_Entropy_FPE}
\end{equation}
Equation~(\ref{Perelman_F_Entropy_FPE}) corresponds to the information-geometric analog of Perelman's entropy functional~\cite{Perelman2002}. The weighting factor \(\sqrt{g} e^{-\Phi}\) is introduced from the probability density mapping, where the ensemble distribution is integrated with the Riemannian volume element.

The temporal evolution of the information metric \(g_{ab}\) can be expressed as a Riemannian gradient flow of this global functional. Taking the functional variational derivative of \(\mathcal{F}\) with respect to the inverse metric yields:
\begin{equation}
\dot{g}_{ab}(t,\theta^{a})
=
- 2 R_{ab}(t,\theta^{a})
- 2 \nabla_{a} \nabla_{b} \Phi(t,\theta^{a})
=
- 2 \frac{\delta \mathcal{F}[g_{ab}(t,\theta^{a})]}
{\delta g^{ab}(t,\theta^{a})} .
\label{FPE_Ricci_Flow_Gradient_Flow}
\end{equation}
Equation~(\ref{FPE_Ricci_Flow_Gradient_Flow}) shows that the dynamics driven by the Fokker-Planck equation can be formulated as a modified Ricci flow that deforms the geometric \(\mathcal{F}\)-entropy landscape. The expression of the metric deformation rate via the functional derivative \(-2\frac{\delta \mathcal{F}}{\delta g^{ab}}\) indicates that the potential \(\Phi \) accounts for the coordinate-dependent components.

\section{From Functional Renormalization Group to Ricci flow}
\label{FRG_RF}

\subsection{A review on the FRG equation}
\label{sec:frg-fp-lift}
Motivated by the probabilistic interpretation of exact renormalization-group flows, in which an effective theory is represented by a scale-dependent probability functional and its RG evolution is formulated as a functional convection--diffusion or Fokker--Planck flow~\cite{Berman2023BayesianRG}, we extend the geometric equivalence between statistical diffusion and the modified Ricci flow established in Section~\ref{FPE_RF} to quantum field theory. 
We therefore define the probability density functional for the statistical distribution of field configurations at a cutoff scale \(\ln\Lambda\):
\bqa && P[\ln \Lambda,\phi(x);\lambda^{a}] = \frac{1}{Z(\ln \Lambda;\lambda^{a})} e^{- S[\ln \Lambda,\phi(x);\lambda^{a}]} . \eqa
Here, $\lambda^a$ denotes the set of coupling constants spanning the theory space, and $Z(\ln \Lambda; \lambda^a)$ represents the partition function serving as the normalization factor obtained by integrating over the field configurations:
\bqa && Z(\ln \Lambda;\lambda^{a}) = \int D \phi(x)\ e^{-S[\ln \Lambda,\phi(x);\lambda^{a}]} . \eqa
The total action $S$ in the exponent is decomposed into a quadratic Gaussian part governing the free propagation of fields and an interaction functional $S_{\mathrm{int}}$ containing the multi-local operators determined by the couplings:
\bqa && S[\ln \Lambda,\phi(x);\lambda^{a}] = \frac{1}{2} \int \frac{d^{d} p}{(2\pi)^{d}} \phi(p) G^{-1}(p^{2}) K_{\Lambda}^{-1}(p^{2}) \phi(-p) + S_{\mathrm{int}}[\ln \Lambda,\phi(x);\lambda^{a}] . \label{RG_Setting} \eqa
In this representation, $G(p^2)$ is the bare propagator, and $K_{\Lambda}(p^2)$ denotes a momentum-dependent cutoff function satisfying the boundary conditions $K_{\Lambda} \to 1$ for $p \ll \Lambda$ and $K_{\Lambda} \to 0$ for $p \gg \Lambda$. This function suppresses high-energy fluctuations to define an effective theory at the scale $\Lambda$. This distribution structure maps the scale-dependent trajectory of the theory onto the geometric framework by treating the RG process as the evolution of a statistical ensemble.

Since the integral of the probability density functional $P$ is conserved under continuous deformations of the energy scale, the probability distribution satisfies the functional normalization invariance condition:
\bqa && \frac{d }{d \ln \Lambda} \int D \phi(x) \ P[\ln \Lambda,\phi(x);\lambda^{a}] = 0 . \eqa
Based on this invariance, the continuous scale evolution of \(P\) governed by the Polchinski exact RG equation \cite{Polchinski1984} can be formulated as an infinite-dimensional Fokker-Planck equation:
\begin{equation}
\begin{aligned}
& \revblue{-}\frac{d }{d \ln \Lambda}
P[\ln \Lambda,\phi(x);\lambda^{a}]\\
&=
\int_{M} d^{d} x \int_{M} d^{d} y
\Big\{
C_{\Lambda}^{Pol.}(x,y)
\frac{\delta^{2} P[\ln \Lambda,\phi(x);\lambda^{a}]}
{\delta \phi(x) \delta \phi(y)}
+
\frac{\delta}{\delta \phi(x)}
\Big(
P[\ln \Lambda,\phi(x);\lambda^{a}]
C_{\Lambda}^{Pol.}(x,y)
\frac{\delta V_{Pol.}[\ln \Lambda,\phi(x)]}{\delta \phi(y)}
\Big)
\Big\}\\
&\equiv
\Delta P[\ln \Lambda,\phi(x);\lambda^{a}]
+
\mbox{div}
\Big(
P[\ln \Lambda,\phi(x);\lambda^{a}]
\mbox{grad}_{C_{\Lambda}^{Pol.}}
V_{Pol.}[\ln \Lambda,\phi(x)]
\Big) .
\end{aligned}
\label{FRG_Eq}
\end{equation}
Equation (\ref{FRG_Eq}) expresses the Polchinski-type renormalization group equation as a statistical diffusion process, embedding the Wilsonian coarse-graining paradigm \cite{Wilson1971, Wegner1973, Wilson1974} within a stochastic framework. 
This interpretation is consistent with the optimal-transport formulation of exact RG flows, in which Polchinski's equation is recast as a Wasserstein gradient flow of a field-theoretic relative entropy~\cite{CotlerRezchikov2023}.
This functional formulation relates to the exact RG flows evaluated via the Wetterich equation or Morris's formulation \cite{Wetterich1993, Morris1994, Ellwanger1994, Berges2002, Pawlowski2007, Kopietz2010, Dupuis2021}. The diffusion kernel (\(C_{\Lambda }^{Pol}\)) and the effective potential (\(V_{Pol}\)) are defined by the scale derivatives of the cutoff function and the free action:
\begin{equation}
\begin{aligned}
C_{\Lambda}^{Pol.}(p^{2})
&=
{\frac{1}{2}}(2\pi)^{d} G(p^{2})
\frac{\partial K_{\Lambda}(p^{2})}{\partial \ln \Lambda} ,\\
V_{Pol.}[\ln \Lambda,\phi(x)]
&=
{
\int \frac{d^{d} p}{(2\pi)^{d}}
\phi(p) G^{-1}(p^{2}) K_{\Lambda}^{-1}(p^{2}) \phi(-p)
} .
\end{aligned}
\label{Cuttoff_Potential}
\end{equation}
With these definitions, the normalized free Gaussian distribution (\(S_{\mathrm{int}}=0\)) satisfies Eq.~\eqref{FRG_Eq}, in agreement with Ref.~\cite{CotlerRezchikov2023}.
To express this functional differential equation, we introduce the following infinite-dimensional operators:
\bea
\Delta &\equiv& \int_{M} d^{d} x \int_{M} d^{d} y~ C_{\Lambda}^{Pol.}(x,y) \frac{\delta^{2} }{\delta \phi(x) \delta \phi(y)}\ ,\\
\mbox{grad}_{C_{\Lambda}^{Pol.}} &\equiv& \int_{M} d^{d} y~ C_{\Lambda}^{Pol.}(x,y) \frac{\delta }{\delta \phi(y)}\ ,\\
\mbox{div} &\equiv& \int_{M} d^{d} x \frac{\delta}{\delta \phi(x)}\ .
\eea
These operators define the Laplacian (\(\Delta \)), gradient (\(\text{grad}\)), and divergence (\(\text{div}\)) in the field configuration space, providing the notation to describe the RG flow in the language of geometric diffusion.

Using the operators defined above, the evolution of the probability density under the RG flow is expressed as a continuity equation in field space:
\bqa
&& \revblue{-}\frac{d }{d \ln \Lambda}
P[\ln \Lambda,\phi(x);\lambda^{a}]
=
- \int_{M} d^{d} x\,
\frac{\delta}{\delta \phi(x)}
\Big(
\Psi[\ln \Lambda,\phi(x);\lambda^{a}]
P[\ln \Lambda,\phi(x);\lambda^{a}]
\Big) ,
\label{FRG_Conservation_Law}
\eqa
where \(\Psi \) represents the probability velocity field in field space, derived from the diffusion and drift terms of the Fokker-Planck equation:
\bqa && \Psi[\ln \Lambda,\phi(x);\lambda^{a}] ~ P[\ln \Lambda,\phi(x);\lambda^{a}] \nn && = - \int_{M} d^{d} y \Big\{ C_{\Lambda}^{Pol.}(x,y) \frac{\delta P[\ln \Lambda,\phi(x);\lambda^{a}]}{\delta \phi(y)} +  P[\ln \Lambda,\phi(x);\lambda^{a}]~ C_{\Lambda}^{Pol.}(x,y) \frac{\delta V_{Pol.}[\ln \Lambda,\phi(x)]}{\delta \phi(y)}  \Big\} . \label{Conserved_Current_FRG} \eqa 
The velocity field defined by Eq. (\ref{Conserved_Current_FRG}) satisfies the conservation of total probability:
\bqa
&& \revblue{-}\frac{d }{d \ln \Lambda}
\int D \phi(x)\ P[\ln \Lambda,\phi(x);\lambda^{a}]
=
- \int D \phi(x)\int_{M} d^{d} x
\frac{\delta}{\delta \phi(x)}
\Big(
\Psi[\ln \Lambda,\phi(x);\lambda^{a}]
P[\ln \Lambda,\phi(x);\lambda^{a}]
\Big)=0 .
\label{Conservation}
\eqa
This velocity field \(\Psi \) can be expressed as the gradient of an effective potential, \(\Sigma\):
 \bqa && \Psi[\ln \Lambda,\phi(x);\lambda^{a}] = \int_{M} d^{d} y~ C_{\Lambda}^{Pol.}(x,y) \frac{\delta \Sigma[\ln \Lambda,\phi(x);\lambda^{a}]}{\delta \phi(y)} \equiv \mbox{grad}_{C_{\Lambda}^{Pol.}} \Sigma[\ln \Lambda,\phi(x);\lambda^{a}]] . \label{Current_Fick_Law} \eqa
The effective potential \(\Sigma \) is the functional relative entropy (or Kullback-Leibler divergence) \cite{Kullback1951, Cover2006, Amari2016} between the total probability density \(P\) and the bare Gaussian weight \(e^{-V_{Pol.}}\):
\bqa && \Sigma[\ln \Lambda,\phi(x);\lambda^{a}] = - \ln \left(\frac{P[\ln \Lambda,\phi(x);\lambda^{a}]}{e^{- V_{Pol.}[\ln \Lambda,\phi(x)]}}\right) = S[\ln \Lambda,\phi(x);\lambda^{a}] - V_{Pol.}[\ln \Lambda,\phi(x)] . \label{Relative_Entropy_FRG} \eqa 
Accordingly, Eqs.~(\ref{FRG_Conservation_Law})--(\ref{Relative_Entropy_FRG}) show that the RG flow follows the gradient of this functional relative entropy in field space. This relative-entropy gradient-flow structure is consistent with the optimal-transport formulation of exact RG flows, in which Polchinski's equation is recast as a Wasserstein gradient flow of a field-theoretic relative entropy~\cite{CotlerRezchikov2023}. It provides the dynamical formulation for the geometric mapping to the Ricci flow discussed in Section~\ref{FPE_RF}.

To extend the geometric equivalence established in Section \ref{FPE_RF} to the non-perturbative domain of the FRG, the statistical drift vector field \(A_{i}(x)\) and the diffusion tensor \(B^{ij}(x)\) are identified with the functional operators of the Polchinski framework via the following algebraic mapping:
\begin{equation}
A^{i}(x)
\longrightarrow
-\int_{M} d^{d}y\,
C_{\Lambda}^{\rm Pol.}(x,y)
\frac{\delta V_{\rm Pol.}[\ln \Lambda,\phi(x)]}{\delta \phi(y)} .
\label{eq:fp-frg-drift-map}
\end{equation}

\begin{equation}
B^{ij}(x)
\longrightarrow
C_{\Lambda}^{\rm Pol.}(x,y) .
\label{eq:fp-frg-diffusion-map}
\end{equation}
Equation~\eqref{eq:fp-frg-drift-map} maps the functional derivative of the effective bare potential \(V_{\rm Pol.}\) in configuration space to the macroscopic drift.
Concurrently, Eq.~\eqref{eq:fp-frg-diffusion-map} identifies the Polchinski kernel \(C_{\Lambda}^{\rm Pol.}\), determined by the scale evolution of the momentum cutoff, with the infinite-dimensional statistical diffusion coefficient.
These relations formulate the FRG scale evolution as a diffusion process in field-configuration space.
Because this mapping involves finite-dimensional state-space variables, spacetime labels, and coupling coordinates at different levels, the Fokker--Planck--FRG correspondence used throughout this section is summarized in {Table~\ref{tab:fp-frg-correspondence} and} Appendix~\ref{app:fp-frg-dictionary}.

\begin{table}[!htb]
\caption{{Core correspondence used to lift the finite-dimensional Fokker--Planck construction of Section~\ref{FPE_RF} to the Polchinski FRG. The state variable on the FRG side is the full field configuration \(\phi(x)\); the spacetime coordinates \(x,y\) are continuous labels corresponding to the finite-dimensional indices \(i,j\). The reference cutoff \(\Lambda_0\) is fixed.}}
\label{tab:fp-frg-correspondence}
\begin{center}
\begingroup
%\color{blue}
\small
\renewcommand{\arraystretch}{1.30}
\begin{tabular}{@{}lcl@{}}
\hline\hline
\textbf{Fokker--Planck formulation}
&&
\textbf{Polchinski FRG formulation} \\
\hline
State-space point \(x=(x^1,\ldots,x^d)\)
& \(\longleftrightarrow\) &
Field configuration \(\phi(x)\) \\
State-space indices \(i,j\)
& \(\longleftrightarrow\) &
Continuous spacetime labels \(x,y\) \\
Time variable \(t\)
& \(\longleftrightarrow\) &
RG time \(\ln(\Lambda_0/\Lambda)\) \\
Statistical parameters \(\theta^a\)
& \(\longleftrightarrow\) &
Coupling coordinates \(\lambda^a\) \\
Probability density \(p(x,t;\theta^a)\)
& \(\longleftrightarrow\) &
Probability functional \(P[\ln\Lambda,\phi;\lambda^a]\) \\
State-space integral \(\int d^dx\)
& \(\longleftrightarrow\) &
Functional integral \(\int D\phi\) \\
State-space derivative \(\partial_i\)
& \(\longleftrightarrow\) &
Functional derivative \(\delta/\delta\phi(x)\) \\
Diffusion tensor \(B^{ij}\)
& \(\longleftrightarrow\) &
Polchinski kernel \(C_{\Lambda}^{\rm Pol.}(x,y)\) \\
Scalar potential \(V\)
& \(\longleftrightarrow\) &
Polchinski potential \(V_{\rm Pol.}\) \\
Drift \(A^i=-B^{ij}\partial_jV\)
& \(\longleftrightarrow\) &
Field-space drift \(-\mbox{grad}_{C_{\Lambda}^{\rm Pol.}}V_{\rm Pol.}\) \\
Probability velocity \(v^i\)
& \(\longleftrightarrow\) &
Field-space probability velocity \(\Psi\) \\
Fisher metric \(g_{ab}\)
& \(\longleftrightarrow\) &
Coupling-space Fisher metric \(G_{ab}\) \\
\hline\hline
\end{tabular}
\endgroup
\end{center}
\end{table}
% Keep Table I before Section III B while allowing text below it.
\FloatBarrier

\subsection{From Fokker-Planck type FRG equation to information metric flow}
\label{sec:frg-metric-lift}

To analyze the information geometry induced by scale transformations in the FRG framework, we define the Fisher information metric \(G_{ab}\) on the coupling-constant manifold associated with the probability functional family \(P[\ln\Lambda,\phi(x);\lambda^a]\):

\bqa && G_{ab}(\ln \Lambda,\lambda^{a}) = \int D \phi(x) P[\ln \Lambda,\phi(x);\lambda^{a}] \partial_{a} \ln P[\ln \Lambda,\phi(x);\lambda^{a}] \partial_{b} \ln P[\ln \Lambda,\phi(x);\lambda^{a}] . \label{FRG_Fisher_Information_Metric} \eqa
Equation (\ref{FRG_Fisher_Information_Metric}) quantifies the statistical distinguishability of the probability functional \(P\) under an infinitesimal variation of the coupling constants \(\lambda ^{a}\), defining a Riemannian structure on the theory space.

When the probability density functional \(P\) evolves according to the infinite-dimensional Fokker-Planck equation, the scale derivative of the information metric \(G_{ab}\) with respect to \(\ln \Lambda\) is given by the flow equation:
\bqa && \revblue{-}\frac{d }{d \ln \Lambda} G_{ab}(\ln \Lambda,\lambda^{a}) = \int D \phi(x) P[\ln \Lambda,\phi(x);\lambda^{a}] \int_{M} d^{d} x \Big[ \Big(\frac{\partial}{\partial \lambda^{a}} \mathcal{V}[\ln \Lambda,\phi(x);\lambda^{a}] \Big) \frac{\delta}{\delta \phi(x)} \Big( \frac{\partial}{\partial \lambda^{b}} \ln P[\ln \Lambda,\phi(x);\lambda^{a}] \Big) \nn && + \Big(\frac{\partial}{\partial \lambda^{b}} \mathcal{V}[\ln \Lambda,\phi(x);\lambda^{a}] \Big) \frac{\delta}{\delta \phi(x)} \Big( \frac{\partial}{\partial \lambda^{a}} \ln P[\ln \Lambda,\phi(x);\lambda^{a}] \Big) \Big] , \label{FRG_Metric_Dynamics} \eqa
where \(\mathcal{V}[\ln \Lambda, \phi(x); \lambda^a]\) is the local effective velocity field determined by the cutoff kernel and interaction terms:
\bqa && \mathcal{V}[\ln \Lambda,\phi(x);\lambda^{a}] = - \int_{M} d^{d} y \Big\{ \Big( C_{\Lambda}^{Pol.}(x,y) \frac{\delta V_{Pol.}[\ln \Lambda,\phi(x)]}{\delta \phi(y)} \Big) \nn && + \frac{1}{P[\ln \Lambda,\phi(x);\lambda^{a}]} C_{\Lambda}^{Pol.}(x,y)\frac{\delta}{\delta \phi(y)} \Big(  P[\ln \Lambda,\phi(x);\lambda^{a}] \Big) \Big\} . \label{FRG_Eff_Force_Field} \eqa 
Equation (\ref{FRG_Eff_Force_Field}) combines the drift and diffusion components of the Polchinski renormalization process. Accordingly, Eq. (\ref{FRG_Metric_Dynamics}) shows that the evolution of the information metric depends on the parametric variations of this effective potential coupled with the informational score functions, extending the finite-dimensional relations established in Section \ref{FPE_RF} to the infinite-dimensional domain.

\newpage
\subsection{From the information metric flow to the Ricci flow}
\label{sec:frg-ricci-lift}

{To give the metric evolution~\eqref{FRG_Metric_Dynamics} a geometric interpretation, we follow the modified Ricci-flow construction of Section~\ref{FPE_RF} and Appendix~A, with \(X_a=-\nabla_a\Phi\). Matching Eq.~\eqref{Modified_Ricci_Flow_1st} to this evolution gives the constraint}

\bqa && -{2}\nabla_{a}\nabla_{b} \Phi  = 2 R_{ab}(\ln \Lambda,\lambda^{a}) \nn && + \int D \phi(x) {P[\ln \Lambda,\phi(x);\lambda^{a}]} \int_{M} d^{d} x\Big[ \Big(\frac{\partial}{\partial \lambda^{a}} \mathcal{V}[\ln \Lambda,\phi(x);\lambda^{a}] \Big) \frac{\delta}{\delta \phi(x)} \Big( \frac{\partial}{\partial \lambda^{b}} \ln P[\ln \Lambda,\phi(x);\lambda^{a}] \Big) \nn && + \Big(\frac{\partial}{\partial \lambda^{b}} \mathcal{V}[\ln \Lambda,\phi(x);\lambda^{a}] \Big) \frac{\delta}{\delta \phi(x)} \Big( \frac{\partial}{\partial \lambda^{a}} \ln P[\ln \Lambda,\phi(x);\lambda^{a}] \Big) \Big] . \label{FRG_To_Determine_Xa} \eqa
{Here \(R_{ab}\) and \(\nabla_a\) are the Ricci tensor and the Levi-Civita covariant derivative associated with the coupling-space Fisher metric \(G_{ab}\).}
This differential constraint defines the potential-driven diffeomorphism that establishes covariance along the Ricci flow trajectory.

\subsection{$\mathcal{F}-$entropy functional}
\label{sec:frg-entropy-lift}

To analyze the relation between the statistical state and its scale-dependent evolution, we extend the thermodynamic free energy functional $\mathcal{E}[P]$ to the infinite-dimensional configuration space within the functional JKO-Wasserstein paradigm~\cite{Jordan1998, Otto2001}:
\begin{equation}
\mathcal{E}[P[\ln \Lambda,\phi(x);\lambda^{a}]] = \int D \phi(x) \, P[\ln \Lambda,\phi(x);\lambda^{a}] \Big( \ln P[\ln \Lambda,\phi(x);\lambda^{a}] + V_{Pol.}[\ln \Lambda,\phi(x)] \Big) ,
\label{Infinite_Free_Energy_Functional}
\end{equation}
where $V_{Pol.}$ is the Polchinski effective potential defined in Eq.~(\ref{Cuttoff_Potential}). Under this formulation, the Polchinski Fokker-Planck equation in Eq.~(\ref{FRG_Eq}) is expressed as a functional gradient flow representation via the functional variation of this RG-flow energy functional $\mathcal{E}[P]$:
\begin{equation}
\revblue{-}\frac{d}{d \ln \Lambda} P[\ln \Lambda,\phi(x);\lambda^{a}] = \int_{M} d^{d}x \int_{M} d^{d}y \, \frac{\delta}{\delta \phi(x)} \left[ P[\ln \Lambda,\phi(x);\lambda^{a} ] C_{\Lambda}^{Pol.}(x,y) \frac{\delta}{\delta \phi(y)} \left( \frac{\delta \mathcal{E}[P[\ln \Lambda,\phi(x);\lambda^{a}]]}{\delta P[\ln \Lambda,\phi(x);\lambda^{a}]} \right) \right] .
\label{FRG_True_Gradient_Flow}
\end{equation}
The formulation of the renormalization group as a functional gradient flow in Eq.~(\ref{FRG_True_Gradient_Flow}) determines the structure of the probability velocity field \(\Psi \) that governs the distribution's motion in field space. In accordance with the relation between the velocity field \(v_{i}\) and the entropy variation evaluated in the finite-dimensional case, the functional velocity field \(\Psi \) is derived as:
\bqa && \Psi[\ln \Lambda,\phi(x);\lambda^{a}] = - \int_{M} d^{d} y ~ C_{\Lambda}^{Pol.}(x,y) \frac{\delta}{\delta \phi(y)} \Bigg\{ \frac{\delta \mathcal{E}[P[\ln \Lambda,\phi(x);\lambda^{a}]]}{\delta P[\ln \Lambda,\phi(x);\lambda^{a}]} \Bigg\} . \label{FRG_Current_W_Entropy} \eqa
Equation~(\ref{FRG_Current_W_Entropy}) defines the mechanism of the FRG scale evolution, where the velocity field \(\Psi \) corresponds to the functional derivative on the \(\mathcal{E}\)-energy landscape weighted by the Polchinski kernel \(C^{Pol.}_{\Lambda}(x, y)\), which operates as a local metric in the configuration space to determine the trajectory of the field distribution.

The rate of this energy dissipation along the renormalization trajectory defines the RG-flow entropy functional, relating to Perelman's energy framework~\cite{Perelman2002}.
\begin{align}
& \mathcal{F}[P[\ln \Lambda,\phi(x);\lambda^{a}]] \equiv {\frac{1}{2}}\frac{d\mathcal{E}[P[\ln \Lambda,\phi(x);\lambda^{a}]]}{d\ln\Lambda} = {\frac{1}{2}} \int D \phi(x) \, \left( \frac{\delta \mathcal{E}[P[\ln \Lambda,\phi(x);\lambda^{a}]]}{\delta P[\ln \Lambda,\phi(x);\lambda^{a}]} \right) \frac{d P[\ln \Lambda,\phi(x);\lambda^{a}]}{d\ln\Lambda} \nonumber \\
&= \int D \phi(x) \, P[\ln \Lambda, \phi(x); \lambda^a] \int_{M} d^{d}x \int_{M} d^{d}y \, \left\{ \frac{1}{2} \left( \frac{\delta \ln P[\ln \Lambda,\phi(x);\lambda^{a}]}{\delta \phi(x)} \right) C_{\Lambda}^{Pol.}(x,y) \left( \frac{\delta \ln P[\ln \Lambda,\phi(x);\lambda^{a}]}{\delta \phi(y)} \right) \right. \nonumber \\
&\left. + \frac{1}{2} \left(\frac{\delta V_{Pol.}[\ln \Lambda,\phi(x)]}{\delta \phi(x)}\right) C_{\Lambda}^{Pol.}(x,y) \left(\frac{\delta V_{Pol.}[\ln \Lambda,\phi(x)]}{\delta \phi(y)}\right) - \frac{\delta}{\delta \phi(x)} \left( C_{\Lambda}^{Pol.}(x,y) \frac{\delta V_{Pol.}[\ln \Lambda,\phi(x)]}{\delta \phi(y)} \right) \right\} .
\label{Lifted_F_Entropy_Perfect}
\end{align} 
In Eq.~(\ref{Lifted_F_Entropy_Perfect}), the first term represents the informational dissipation from stochastic diffusion, and the remaining terms account for the background potential landscape determined by the cutoff kernel.

Taking the second-order variation of this \(\mathcal{F}\)-entropy with respect to the coupling coordinates \(\lambda^{a}\) and \(\lambda^{b}\) gives the functional analogue of the finite-dimensional residual decomposition:
\bqa
&&
{-}\frac{d}{d\ln\Lambda}G_{ab}(\ln\Lambda,\lambda^{a})
=
-2\frac{\partial^{2}\mathcal{F}\Big(P[\ln\Lambda,\phi(x);\lambda^{a}]\Big)}
{\partial\lambda^{a}\partial\lambda^{b}}
\nn
&&
+2\int D\phi(x)\,
\left[
\frac{\partial^{2}P[\ln\Lambda,\phi(x);\lambda^{a}]}
{\partial\lambda^{a}\partial\lambda^{b}}
\mathcal{H}_{\rm functional}+P[\ln\Lambda,\phi(x);\lambda^{a}]
\mathcal{R}^{\rm functional}_{ab}
\right].
\label{FRG_Fisher_Information_Gradient_Flow}
\eqa
The integral term in Eq.~\eqref{FRG_Fisher_Information_Gradient_Flow} defines the functional residual sector.
Here \(\mathcal{H}_{\rm functional}\) is the functional potential-constraint block, and \(\mathcal{R}^{\rm functional}_{ab}\) is the corresponding parametric remnant.
They are the field-theoretic lifts of the finite-dimensional residuals derived in Appendix~B, under the dictionary summarized in Appendix~C.
Thus Eq.~\eqref{FRG_Fisher_Information_Gradient_Flow} should be read as an off-shell functional metric-flow decomposition: the first term is the direct parametric Hessian of \(\mathcal{F}[P]\), while the second term records the residual mismatch removed in the compatible zero-current sector.
The explicit functional forms of these residuals are not displayed, because not inspiring.

To convert this metric-flow relation into a curvature relation on the coupling-constant manifold, we compare Eq.~\eqref{FRG_Fisher_Information_Gradient_Flow} with {the constraint~\eqref{FRG_To_Determine_Xa} obtained from Eq.~\eqref{Modified_Ricci_Flow_1st}}. Using the same potential convention as above, \(X_a=-\nabla_a\Phi\), the diffeomorphism contribution becomes \(\nabla_aX_b+\nabla_bX_a=-2\nabla_a\nabla_b\Phi\).
Equating the two expressions for \(\revblue{-}dG_{ab}/d\ln\Lambda\) and dividing by \(-2\) then gives the residual-inclusive relation between the Ricci curvature and the entropic variation:

\bqa
&&
R_{ab}(\ln\Lambda,\lambda^{a})
+
\nabla_{a}\nabla_{b}\Phi(\ln\Lambda,\lambda^{a})
=
\frac{\partial^{2}
\mathcal{F}\Big(P[\ln\Lambda,\phi(x);\lambda^{a}]\Big)}
{\partial\lambda^{a}\partial\lambda^{b}}
\nn
&&
-
\int D\phi(x)\,
\left[
\frac{\partial^{2}P[\ln\Lambda,\phi(x);\lambda^{a}]}
{\partial\lambda^{a}\partial\lambda^{b}}
\mathcal{H}_{\rm functional}
+
P[\ln\Lambda,\phi(x);\lambda^{a}]
\mathcal{R}^{\rm functional}_{ab}
\right].
\label{FRG_Constraint_Eq}
\eqa

The residual integral in Eq.~\eqref{FRG_Constraint_Eq} is removed by restricting the functional flow to the compatible zero-current sector at fixed RG scale.
In this sector, \(P[\phi]\propto e^{-V_{\rm Pol.}[\ln\Lambda,\phi]},\) and the field-space probability current vanishes.
The finite-dimensional cancellation derived in Appendix~B then lifts directly to field-configuration space: \(\mathcal{H}_{\rm functional}\) vanishes pointwise, while the weighted remnant \(P\mathcal{R}^{\rm functional}_{ab}\) becomes a functional total divergence.
Thus the residual integral in Eq.~\eqref{FRG_Constraint_Eq} gives no contribution under the same functional boundary conditions used in the integrations by parts.
Equivalently, this compatible zero-current sector is the Wasserstein gradient-transport sector of the functional probability flow, with \(C_{\Lambda}^{\rm Pol.}(x,y)\) serving as the field-space mobility kernel~\cite{Polchinski1984,Jordan1998,Otto2001,Villani2009}.
Equation~\eqref{FRG_Constraint_Eq} therefore reduces to the residual-free curvature--Hessian balance governing the covariant geometric flow.

This reduced relation is the FRG analogue of the modified steady Ricci-soliton equation.
The scalar potential \(\Phi\) supplies the diffeomorphism component required for covariance under reparametrizations of the coupling coordinates, while the parametric Hessian of \(\mathcal{F}[P]\) supplies the entropic driving term.
Accordingly, RG fixed points are formulated as geometric equilibria in which the intrinsic Ricci curvature of the coupling-constant manifold is balanced by the Hessian contribution generated by \(\Phi\).

Based on this relation, the global geometric $\mathcal{F}$-entropy functional on the coupling constant manifold is defined as:
\bqa && \mathcal{F}[G_{ab}(\ln \Lambda,\lambda^{a})] = \int d^{N} \lambda \sqrt{G(\ln \Lambda,\lambda^{a})} e^{- \Phi(\ln \Lambda,\lambda^{a})} \Big\{ R(\ln \Lambda,\lambda^{a}) + G^{ab}(\ln \Lambda,\lambda^{a}) \partial_{a} \Phi(\ln \Lambda,\lambda^{a}) \partial_{b} \Phi(\ln \Lambda,\lambda^{a}) \Big\} . \label{FRG_W_Entropy_Functional_Perellman} \eqa
Equation~(\ref{FRG_W_Entropy_Functional_Perellman}) corresponds to the information-geometric analog of Perelman's entropy functional~\cite{Perelman2002}. The weighting factor $\sqrt{G} e^{-\Phi}$ within the parametric volume element is derived from the information-geometric mapping, allowing the functional to act as a Lyapunov potential for the renormalization group flow.

The scale-dependent evolution of the information metric $G_{ab}$ under the FRG flow can be expressed as a Riemannian gradient flow of this global $\mathcal{F}$-entropy functional:
\bqa && {-}\frac{d }{d \ln \Lambda} G_{ab}(\ln \Lambda,\lambda^{a}) = - 2 R_{ab}(\ln \Lambda,\lambda^{a}) - 2 \nabla_{a} \nabla_{b} \Phi(\ln \Lambda,\lambda^{a}) = - 2 \frac{\delta \mathcal{F}[G_{ab}(\ln \Lambda,\lambda^{a})]}{\delta G^{ab}(\ln \Lambda,\lambda^{a})} . \label{FRG_Ricci_Flow_Gradient_Flow} \eqa
The formulation of the metric deformation rate via the variational derivative $-2 \frac{\delta \mathcal{F}}{\delta G^{ab}}$ shows that the non-perturbative RG process corresponds to a geometric gradient flow, where the potential \(\Phi\) accounts for the coordinate-dependent components to ensure the covariance of the flow with respect to the intrinsic curvature of the parameter space.

%\begingroup
%\color{blue}

\subsection{{Example: a weakly interacting scalar field}}
\label{sec:frg-phi4-example}

We compare field distributions of \(\phi^4\) theories with slightly different quadratic and quartic couplings. The Fisher metric \(G_{ab}\) in Eq.~\eqref{FRG_Fisher_Information_Metric} measures their statistical distinguishability, and Eq.~\eqref{FRG_Metric_Dynamics} determines its change under Polchinski RG evolution. Combining this metric change with the curvature of the same coupling manifold in Eq.~\eqref{FRG_Ricci_Flow_Gradient_Flow} gives the explicit tensor that the information potential \(\Phi\) must reproduce as its covariant Hessian.

The two couplings define a two-dimensional manifold, independently of the spacetime dimension \(d\). This is a minimal setting for studying changes in statistical distinguishability together with nontrivial intrinsic curvature, since one-dimensional manifolds have vanishing intrinsic curvature.

Consider the regulated scalar theory of Eq.~\eqref{RG_Setting} in a finite periodic volume \(\Omega\), with initial action
\begin{equation}
\begin{aligned}
S_{\Lambda_0}[\phi;r,u]
&=S_{0,\Lambda_0}[\phi]+rO_r[\phi]+uO_u[\phi],\\
O_r[\phi]&=\frac12\int_M d^dx\,\phi(x)^2,
\qquad
O_u[\phi]=\frac{1}{4!}\int_M d^dx\,\phi(x)^4.
\end{aligned}
\label{eq:phi4-initial-action}
\end{equation}
Here \(S_{0,\Lambda_0}\) is the quadratic part of Eq.~\eqref{RG_Setting}, \(r\) shifts its mass term, and \(u>0\) is a weak quartic coupling. We assume a positive Gaussian quadratic form. The initial distributions \(P_0[\phi;r,u]=Z^{-1}e^{-S_{\Lambda_0}}\), with expectations \(\langle\cdots\rangle_0\), evolve under Eq.~\eqref{FRG_Eq} with a common kernel \(C_\Lambda^{\rm Pol.}\) and drift potential \(V_{\rm Pol.}\), both independent of \(r,u\) at each RG scale.

As in the interacting scalar-field example in Sec.~V~B of Ref.~\cite{CotlerRezchikov2023}, we evaluate the RG response at the initial cutoff. We write \(\dot G_{ab}=\left.\partial_\tau G_{ab}\right|_{\tau=0}\), where \(\tau=\ln(\Lambda_0/\Lambda)\) increases as the cutoff decreases. The labels \((r,u)\) distinguish the initial distributions and are held fixed during this evolution.

The scores \(\partial_a\ln P_0=-\delta O_a\), with \(\delta O_a=O_a-\langle O_a\rangle_0\), express fluctuations of the coupling observables. Equation~\eqref{FRG_Fisher_Information_Metric} then gives
\begin{equation}
\begin{aligned}
G_{ab}(\Lambda_0)
&=\langle\delta O_a\delta O_b\rangle_0
=\partial_a\partial_b\ln Z,
\qquad a,b\in\{r,u\},\\
G_{rr}(\Lambda_0)
&=\frac14\int_M d^dx\,d^dy\,
\langle\phi(x)^2\phi(y)^2\rangle_{0,c}
=-\partial_r\langle O_r\rangle_0.
\end{aligned}
\label{eq:phi4-score-metric}
\end{equation}
The subscript \(c\) denotes a connected correlation, \(\langle\phi(x)^2\phi(y)^2\rangle_{0,c}=\langle\phi(x)^2\phi(y)^2\rangle_0-\langle\phi(x)^2\rangle_0\langle\phi(y)^2\rangle_0\). Thus \(G_{rr}\) and \(G_{uu}\) are the variances of the quadratic and quartic observables, while \(G_{ru}\) is their covariance~\cite{Floerchinger2024}. At fixed \(u\), \(ds^2=G_{rr}(dr)^2\) identifies the mass susceptibility as the measure of distinguishability between nearby quadratic couplings.

Expanding about \(u=0\), we use the Gaussian propagator \(D(p)=[(D_{\rm bare}(p)K_{\Lambda_0}(p^2))^{-1}+r]^{-1}\). Here \(D_{\rm bare}(p)\equiv G(p^2)\) is the fixed massive bare propagator in Eq.~\eqref{RG_Setting}; \(D\) describes field correlations, while \(G_{ab}\) measures distinguishability in coupling space.

Writing \(\langle\cdots\rangle_{\rm G}\) for the Gaussian average at \(u=0\), we have \(D(x-y)=\langle\phi(x)\phi(y)\rangle_{\rm G}\). Wick contraction then gives \(\langle\phi(x)^2\phi(y)^2\rangle_{{\rm G},c}=2D(x-y)^2\) for the connected correlation in Eq.~\eqref{eq:phi4-score-metric}.

To evaluate \(G_{rr}\), we use the propagator integrals \(I_n=\int d^dp\,D(p)^n/(2\pi)^d\), understood as normalized sums over regulated modes. With \(Z_{\rm G}=Z|_{u=0}\), differentiating \(\ln Z=\ln Z_{\rm G}-u\Omega I_1^2/8+O(u^2)\) twice with respect to \(r\), as prescribed by Eq.~\eqref{eq:phi4-score-metric}, and using \(\partial_r I_n=-nI_{n+1}\) gives
\begin{equation}
\frac{G_{rr}(\Lambda_0)}{\Omega}
=\frac12 I_2-\frac{u}{2}I_1I_3-\frac{u}{4}I_2^2+O(u^2).
\label{eq:phi4-metric-perturbative}
\end{equation}
The Gaussian term and two interaction corrections provide a perturbative benchmark: the tadpole and connected four-point contributions reduce the susceptibility at fixed cutoff and bare \(r\).

The curvature requires the complete \((r,u)\) metric. We evaluate the metric and curvature at the Gaussian boundary \(u\to0^+\), where the required correlations can be computed explicitly by Wick contraction. Coupling derivatives are taken before the limit so that variations in the quartic coupling are retained. The component \(G_{uu}\) also involves the spatial integral \(J_4=\int_M d^dx\,D(x)^4\). The covariances in Eq.~\eqref{eq:phi4-score-metric} then give
\begin{equation}
\left.G_{ab}\right|_{u=0}
=\Omega\begin{pmatrix}
I_2/2 & I_1I_2/4\\
I_1I_2/4 & I_1^2I_2/8+J_4/24
\end{pmatrix},
\qquad
\left.\det G\right|_{u=0}=\frac{\Omega^2 I_2J_4}{48}>0,
\label{eq:phi4-full-gaussian-metric}
\end{equation}
The positive determinant shows that both coupling directions remain distinguishable at \(u=0\). Second and third \(u\)-derivatives of \(\ln Z\) are computed from Gaussian cumulants.

The initial RG change follows from Eq.~\eqref{FRG_Metric_Dynamics} (see Appendix~\ref{appD:initial-metric-deformation} for the derivation). Since the kernel and drift potential are independent of \(r,u\), coupling derivatives of the velocity~\eqref{Conserved_Current_FRG} act only on \(\ln P_0\). Inserting the scores \(\partial_a\ln P_0=-\delta O_a\), together with \(\delta O_r/\delta\phi(x)=\phi(x)\) and \(\delta O_u/\delta\phi(x)=\phi(x)^3/6\), gives
\begin{equation}
\begin{aligned}
\dot G_{ab}
&=-2\int_M d^dx\int_M d^dy\,
C_{\Lambda_0}^{\rm Pol.}(x,y)
\left\langle
\frac{\delta O_a}{\delta\phi(x)}
\frac{\delta O_b}{\delta\phi(y)}
\right\rangle_0,\\
\dot G_{rr}
&=-2\operatorname{Tr}(CD)
+uI_1\operatorname{Tr}(CD^2)+O(u^2).
\end{aligned}
\label{eq:phi4-initial-metric-flow}
\end{equation}
The second line is the first-order expansion about \(u=0\) at fixed \(r\), with \(D\) evaluated in the Gaussian theory. Here \(C=C_{\Lambda_0}^{\rm Pol.}\), traces run over regulated field modes, and \(D^2\) denotes operator composition. For a symmetric kernel \(C\) with nonnegative eigenvalues, the first line of Eq.~\eqref{eq:phi4-initial-metric-flow} gives \(\dot G_{ab}\,d\lambda^a d\lambda^b\le0\) for every fixed infinitesimal separation \(d\lambda=(dr,du)\) (see Eq.~\eqref{appD:distance-contraction}). Thus \(ds^2=G_{ab}d\lambda^a d\lambda^b\), the squared statistical distance between neighboring distributions, cannot increase in the first RG step, consistent with the contraction of distinguishability under coarse graining~\cite{BenyOsborne2015PRA,BenyOsborne2015NJP}. The positive \(O(u)\) correction makes \(\dot G_{rr}\) less negative: weak repulsion reduces the initial loss of distinguishability along the mass direction.

At \(u=0\), all three metric components evolve according to the same identity. The RG kernel enters through the integrals \(T_C=\operatorname{Tr}(CD)\) and \(U_C=\int_M d^dx\,d^dy\,C(x,y)D(x-y)^3\), so Eq.~\eqref{eq:phi4-initial-metric-flow} gives
\begin{equation}
\left.\dot G_{ab}\right|_{u=0}
=-2\begin{pmatrix}
T_C&I_1T_C/2\\
I_1T_C/2&I_1^2T_C/4+U_C/6
\end{pmatrix}.
\label{eq:phi4-full-gaussian-flow}
\end{equation}
We now compute the curvature of the coupling manifold, the remaining input to Eq.~\eqref{FRG_Ricci_Flow_Gradient_Flow}. In two dimensions, the Ricci tensor is determined by the scalar curvature and the metric. With \(W=\ln Z\), the Hessian structure \(G_{ab}=W_{ab}\equiv\partial_a\partial_b W\) further simplifies the general connection and curvature formulas~\eqref{Connection_Torsion_Free}--\eqref{Ricci_Curvature_Original}, yielding an expression involving only second and third coupling derivatives (see Appendix~\ref{app:phi4-example-details} for the derivation):
\begin{equation}
R=-\frac{1}{2(\det G)^2}
\begin{vmatrix}
W_{rr}&W_{ru}&W_{uu}\\
W_{rrr}&W_{rru}&W_{ruu}\\
W_{rru}&W_{ruu}&W_{uuu}
\end{vmatrix},
\qquad R_{ab}=\frac R2G_{ab}.
\label{eq:phi4-hessian-curvature}
\end{equation}
Here \(R\) is the scalar curvature, and subscripts on \(W\) denote coupling derivatives. The \(3\times3\) determinant contains the three independent metric components \(G_{rr},G_{ru},G_{uu}\) in its first row and their \(r\)- and \(u\)-derivatives in the next two rows.

At \(u=0\), \(W_{abc}=-\langle\delta O_a\delta O_b\delta O_c\rangle_{\rm G}\). The mixed quadratic--quartic and purely quartic correlations entering these derivatives involve the spatial integrals \(J_5=\int_M d^dx\,d^dy\,D(x)^3D(y)D(x-y)\) and \(J_6=\int_M d^dx\,d^dy\,D(x)^2D(y)^2D(x-y)^2\). Substituting the derivatives in Eq.~\eqref{appD:gaussian-third-derivatives} gives
\begin{equation}
\left.R\right|_{u=0}
=\frac{1}{\Omega}\left[
\frac{3I_2^2}{J_4}
-\frac{8I_3J_5}{I_2^2J_4}
+\frac{16J_5^2}{I_2J_4^2}
-\frac{18I_2J_6}{J_4^2}
\right],
\label{eq:phi4-gaussian-curvature}
\end{equation}
which, together with the metric~\eqref{eq:phi4-full-gaussian-metric}, fixes the Ricci tensor \(R_{ab}=RG_{ab}/2\) at \(u=0\).

Thus, the metric and its curvature are determined by second and third connected cumulants of the same coupling observables. This suggests investigating whether the Callan--Symanzik evolution of these correlations~\cite{Callan1970,Symanzik1970} can be organized into a modified Ricci flow with a dilaton-like scalar potential~\cite{Tseytlin2007}. Establishing this connection could give \(\Phi\) a field-theoretic characterization through the beta functions and mixing of composite operators.

For the initial RG step considered here, the calculated metric change and curvature determine the tensor required of \(\Phi\) by the Ricci-flow relation~\eqref{FRG_Ricci_Flow_Gradient_Flow}. In the entropy identity~\eqref{FRG_Fisher_Information_Gradient_Flow}, we write the residual contribution as \(2\mathcal I_{ab}\), evaluated at \(P_0\). With \(\mathcal F[P_0]\) defined in Eq.~\eqref{Lifted_F_Entropy_Perfect}, the two relations give
\begin{equation}
\nabla_a\nabla_b\Phi
=-\frac12\dot G_{ab}-R_{ab}
=\partial_a\partial_b\mathcal F[P_0]-\mathcal I_{ab}-R_{ab}.
\label{eq:phi4-required-hessian}
\end{equation}
The residual can be evaluated from the same initial correlations. Its quadratic component is
\begin{equation}
\left.\mathcal I_{rr}\right|_{u=0}
=3T_C-8r\operatorname{Tr}(CD^2)+4r^2\operatorname{Tr}(CD^3).
\label{eq:phi4-gaussian-residual-rr}
\end{equation}
Even at \(r=u=0\), Eq.~\eqref{eq:phi4-gaussian-residual-rr} gives \(\mathcal I_{rr}=3T_C\), which is generally nonzero: the free Gaussian reference evolves under the Polchinski flow. Appendix~\ref{appD:gaussian-residual} gives the derivation and all three residual components.

Substituting the computed metric~\eqref{eq:phi4-full-gaussian-metric}, deformation~\eqref{eq:phi4-full-gaussian-flow}, and curvature~\eqref{eq:phi4-gaussian-curvature} into the first equality of Eq.~\eqref{eq:phi4-required-hessian} yields the explicit tensor
\begin{equation}
\begin{aligned}
\left[-\frac12\dot G_{ab}-R_{ab}\right]_{u=0}
={}&\left(T_C-\frac{\Omega R I_2}{4}\right)
\begin{pmatrix}
1&I_1/2\\
I_1/2&I_1^2/4
\end{pmatrix}\\
&+\left(\frac{U_C}{6}-\frac{\Omega R J_4}{48}\right)
\begin{pmatrix}
0&0\\
0&1
\end{pmatrix},
\end{aligned}
\label{eq:phi4-explicit-ricci-balance}
\end{equation}
with all quantities evaluated at \(u=0\). The first term contains the quadratic direction and its overlap with the quartic one; the second contributes only to the \(uu\) component. Equation~\eqref{eq:phi4-explicit-ricci-balance} is the explicit result of applying Sections~\ref{sec:frg-metric-lift}--\ref{sec:frg-entropy-lift} to the two-coupling \(\phi^4\) family. At the initial RG scale and Gaussian boundary, the Polchinski kernel and field correlations specify all three covariant Hessian components required of \(\Phi\). This tensor must accompany the Ricci term to reproduce the calculated RG change of statistical distinguishability.

%\par
%\endgroup

\section{Conclusion}

The analysis developed here formulates the Polchinski Functional Renormalization Group as a functional JKO--Wasserstein gradient flow of field distributions driven by a \lq\lq thermodynamic\rq\rq\ free energy. Within this formulation, the scale derivative of the Fisher information metric on the coupling-constant manifold decomposes into the intrinsic Ricci curvature $R_{ab}$ and the covariant Hessian $\nabla_a\nabla_b\Phi$ of an emergent scalar information potential, yielding a modified Ricci flow of theory space. The potential $\Phi$ generates the diffeomorphism component associated with reparametrizations of the coupling coordinates, so that the combination $R_{ab}+\nabla_a\nabla_b\Phi$ governs the covariant geometric evolution of the information metric.
{The characterization of \(\Phi\) nevertheless remains implicit, and the existence of a solution satisfying the full tensor equation~\eqref{Modified_Ricci_Flow_1st} with suitable boundary conditions remains open. A direct physical interpretation beyond its geometric role also remains to be established.}

The same metric evolution admits an entropic formulation through the RG-flow entropy defined by the continuous scale-dissipation rate of the free energy. Its parametric Hessian governs the scale dependence of the Fisher metric, relating the integration of short-distance fluctuations to changes in the statistical distinguishability of neighboring effective theories. This establishes the variational connection between the JKO--Wasserstein formulation of the FRG and the modified Ricci-flow geometry of the coupling manifold.

The stationary limit of this geometric evolution provides a representation of non-perturbative FRG fixed points. When the beta functions vanish and the induced Fisher metric becomes stationary, the flow reduces to the steady Ricci soliton condition $R_{ab}+\nabla_a\nabla_b\Phi=0$. At such fixed points, the intrinsic Ricci curvature of the coupling manifold is balanced by the covariant Hessian of the information potential, providing a geometric characterization of scale-invariant critical behavior and, where applicable, conformal invariance. This perspective connects quantum field theory with geometric evolution and offers a framework for investigating the geometric stability and universality of renormalization-group fixed points.

In this sense, the Ricci-flow formulation turns the geometric characterization of RG fixed points into a possible bridge from RG stability to spectral universality.
Along this bridge, the FRG flow on the coupling-constant manifold and Dyson Brownian motion on the spectral manifold may be regarded as dual modified Ricci-flow descriptions, related by a parameter-to-spectrum pullback map.
If this pullback geometry remains non-degenerate in chaotic regimes, soliton rigidity may then constrain the spectral flow toward the Random-Matrix-Theory universality class.

\begin{acknowledgments}
K.-S. K. was supported by the Ministry of Education, Science, and Technology (Grant No. RS-2024-00337134) of the National Research Foundation of Korea (NRF).
J. M. B. acknowledges support from the National Research Foundation of Korea (NRF) (No. RS2022-NR072365). {ChatGPT 6 (OpenAI), Claude Opus 5.5 (Anthropic), and Gemini (Google) were used to assist with selected derivations and to check calculations. The authors reviewed the results and take full responsibility for the content of the manuscript.}
\end{acknowledgments}

\appendix

\section{Fisher-metric deformation and the diffeomorphism constraint}

To analyze the geometric interpretation of the information metric evolution derived in Eq. (\ref{Metric_Eq_1st}), the flow is decomposed into a Ricci curvature flow and a diffeomorphism component:
\bqa
&& \dot{g}_{ab}(t,\theta^{a})
=
-2R_{ab}(t,\theta^{a})
+\nabla_{a}X_{b}(t,\theta^{a})
+\nabla_{b}X_{a}(t,\theta^{a}) ,
\label{appA:Modified_Ricci_Flow_1st}
\eqa
where \(R_{ab}\) is the intrinsic Ricci curvature tensor of the information space. Comparing Eq. (\ref{appA:Modified_Ricci_Flow_1st}) with Eq. (\ref{Metric_Eq_1st}) suggests the parameter-space field \(X_{a}\) as:
\bqa
&& X_{a}(t,\theta^{a})
=
\int d^{d} x\,
p(x,t;\theta^{a}) v^{i}(x,t;\theta^{a})
\partial_{i}\Big(\partial_{a}\ln p(x,t;\theta^{a})\Big) .
\label{Diffeo_Xa}
\eqa
The covariant derivative of this field, \(\nabla_{a} X_{b}\), is evaluated with respect to the Riemannian connection of the underlying information space:
\bqa
&& \nabla_{a} X_{b}(t,\theta^{a})
=
\partial_{a} X_{b}(t,\theta^{a})
-\Gamma_{ab}^{c}(t,\theta^{a}) X_{c}(t,\theta^{a}) .
\eqa
The Christoffel symbols \(\Gamma _{abc}\) of the torsion-free Levi-Civita connection on the information manifold are derived from the Fisher metric and expressed via the variations of the probability distribution \(p\):
\bqa
&& \Gamma_{abc}(t,\theta^{a})
=
\frac{1}{2}
\Big(
\partial_{a}g_{bc}(t,\theta^{a})
+\partial_{b}g_{ac}(t,\theta^{a})
-\partial_{c}g_{ab}(t,\theta^{a})
\Big)
\nn
&& =
\int d^{d}x\,p(t,x;\theta^{a})
\Big\{
\Big(\partial_{a}\partial_{b}\ln p(t,x;\theta^{a})\Big)
\Big(\partial_{c}\ln p(t,x;\theta^{a})\Big)
\nn
&& \quad+
\frac{1}{2}
\Big(\partial_{a}\ln p(t,x;\theta^{a})\Big)
\Big(\partial_{b}\ln p(t,x;\theta^{a})\Big)
\Big(\partial_{c}\ln p(t,x;\theta^{a})\Big)
\Big\} .
\label{Connection_Torsion_Free}
\eqa

When the field \(X_{a}\) is defined as in Eq. (\ref{Diffeo_Xa}), the relation between Eq. (\ref{appA:Modified_Ricci_Flow_1st}) and Eq. (\ref{Metric_Eq_1st}) determines the statistical curvature field, denoted as $R_{ab}^{\text{naive}}(t,\theta^{a})$. However, this expression does not satisfy the tensor transformation law of Riemannian geometry, resulting in a geometric mismatch:
\bqa
&& R_{ab}^{\text{naive}}(t,\theta^{a})
\not=
\partial_{c}\Gamma_{ab}^{c}(t,\theta^{a})
-\partial_{b}\Gamma_{ac}^{c}(t,\theta^{a})
+\Gamma_{cd}^{c}(t,\theta^{a})\Gamma_{ab}^{d}(t,\theta^{a})
-\Gamma_{bd}^{c}(t,\theta^{a})\Gamma_{ac}^{d}(t,\theta^{a})
=
R_{ab}(t,\theta^{a}) .
\label{Ricci_Curvature_Original}
\eqa
The inequality in Eq. (\ref{Ricci_Curvature_Original}) identifies the discrepancy where the naive curvature calculation deviates from the tensorial property of the standard Ricci tensor.

Using the coordinate-invariant tensor structure, the governing equation for the field \(X_{a}\) required to equate the statistical evolution in Eq. (\ref{Metric_Eq_1st}) with the modified Ricci flow in Eq. (\ref{appA:Modified_Ricci_Flow_1st}) is derived as:
\bqa
&& \nabla_{a}X_{b}(t,\theta^{a})
+\nabla_{b}X_{a}(t,\theta^{a})
=
2R_{ab}(t,\theta^{a})
\nn
&& \quad+
\int d^{d}x\,
{p(x,t;\theta^{a})}
\Big[
\Big(\partial_{a}v^{i}(x,t;\theta^{a})\Big)
\partial_{i}\Big(\partial_{b}\ln p(x,t;\theta^{a})\Big)
+
\Big(\partial_{b}v^{i}(x,t;\theta^{a})\Big)
\partial_{i}\Big(\partial_{a}\ln p(x,t;\theta^{a})\Big)
\Big] .
\label{To_Determine_Xa}
\eqa
Equation (\ref{To_Determine_Xa}) relates the geometric curvature of the information space to the dynamical velocity fields of the statistical current.

\section{Residual terms, Wasserstein compatibility, and zero-current cancellation}
\label{app:residual-cancellation}

In this appendix, we derive the residual sector that appears when the ordinary parametric Hessian of the statistical \(\mathcal{F}\)-entropy is compared with the Fisher-metric deformation generated by the Fokker--Planck probability current. This comparison produces the potential-constraint block \(\mathcal{H}\) and the weighted parametric remnant \(p\mathcal{R}_{ab}\). These terms measure the obstruction to identifying the direct parametric Hessian with the Fisher--Wasserstein Hessian matching in the constant-mobility sector. We then show that, on the stationary zero-current profile, \(\mathcal{H}\) vanishes pointwise while \(p\mathcal{R}_{ab}\) reduces to a total divergence, so that the integrated residual sector gives no contribution under the same boundary conditions used in the integrations by parts. This finite-dimensional cancellation provides the prototype for the functional FRG residual cancellation used in Sec.~{\ref{sec:frg-entropy-lift}}, where the Polchinski kernel plays the role of a fixed field-space mobility kernel at each RG scale and the corresponding lift is summarized in {Table~\ref{tab:fp-frg-correspondence} and} Appendix~\ref{app:fp-frg-dictionary}.

We begin with the brute-force second parameter derivative of the statistical \(\mathcal{F}\)-entropy:
\begin{align}
\partial_a\partial_b \mathcal{F}
=
\int_M d^dx\,\Bigg[
&
(\partial_a p)
\left(
\frac{1}{2}(\partial_b B^{ij})\partial_i\ln p\,\partial_j\ln p
+
\partial_b\widetilde V
\right)
\nonumber\\
&+
p
\left(
\frac{1}{2}(\partial_a\partial_b B^{ij})\partial_i\ln p\,\partial_j\ln p
+
(\partial_b B^{ij})\partial_i(\partial_a\ln p)\,\partial_j\ln p
+
\partial_a\partial_b\widetilde V
\right)
\nonumber\\
&+
(\partial_a\partial_b p)
\left(
\widetilde V
-
\partial_i(B^{ij}\partial_j\ln p)
-
\frac{1}{2}B^{ij}\partial_i\ln p\,\partial_j\ln p
\right)
\nonumber\\
&+
(\partial_b p)
\left(
\partial_a\widetilde V
-
\partial_i\left((\partial_a B^{ij})\partial_j\ln p\right)
-
\partial_i\left(B^{ij}\partial_j(\partial_a\ln p)\right)
\right.
\nonumber\\
&\hspace{2.4cm}
\left.
-
\frac{1}{2}(\partial_a B^{ij})\partial_i\ln p\,\partial_j\ln p
-
B^{ij}\partial_i(\partial_a\ln p)\,\partial_j\ln p
\right)
\Bigg].
\label{appB:bruteforce-hessian}
\end{align}

This expression can be reorganized by comparing it with the Fisher-metric flow generated by the probability-current velocity \(v^i\). The result is
\begin{align}
\partial_a\partial_b \mathcal{F}
=
-\frac{1}{2}\dot g_{ab}
+
\int_M d^dx\,
\left[
(\partial_a\partial_b p)\mathcal{H}
+
p\mathcal{R}_{ab}
\right].
\label{appB:hessian-residual-decomposition}
\end{align}
Here
\begin{equation}
\mathcal{H}[p]
=
\widetilde V
-
\partial_i(B^{ij}\partial_j\ln p)
-
\frac{1}{2}B^{ij}
\partial_i\ln p\,\partial_j\ln p ,
\label{appB:H-definition}
\end{equation}
and the weighted remnant \(p\mathcal{R}_{ab}\) is
\begin{align}
p\mathcal{R}_{ab}
={}&
(\partial_a p)
\biggl[
\frac{1}{2}(\partial_bB^{ij})
\partial_i\ln p\,\partial_j\ln p
+
\partial_b\widetilde V
\biggr]
\nonumber\\
&+
p
\biggl[
\frac{1}{2}(\partial_a\partial_bB^{ij})
\partial_i\ln p\,\partial_j\ln p
+
(\partial_bB^{ij})
\partial_i(\partial_a\ln p)\,\partial_j\ln p
+
\partial_a\partial_b\widetilde V
\biggr]
\nonumber\\
&+
(\partial_b p)
\biggl[
\partial_a\widetilde V
-
\partial_i\bigl((\partial_aB^{ij})\partial_j\ln p\bigr)
-
\partial_i\bigl(B^{ij}\partial_j(\partial_a\ln p)\bigr)
\nonumber\\
&\hspace{3.4cm}
-
\frac{1}{2}(\partial_aB^{ij})
\partial_i\ln p\,\partial_j\ln p
-
B^{ij}\partial_i(\partial_a\ln p)\,\partial_j\ln p
\biggr]
\nonumber\\
&-
\frac{1}{2}p
\biggl[
(\partial_bB^{ij})\partial_j(\ln p+V)\partial_i(\partial_a\ln p)
+
(\partial_aB^{ij})\partial_j(\ln p+V)\partial_i(\partial_b\ln p)
\nonumber\\
&\hspace{2.4cm}
+
B^{ij}\partial_j(\partial_b\ln p+\partial_bV)\partial_i(\partial_a\ln p)
+
B^{ij}\partial_j(\partial_a\ln p+\partial_aV)\partial_i(\partial_b\ln p)
\biggr].
\label{appB:R-definition}
\end{align}
Equation~\eqref{appB:R-definition} should be understood as the off-shell algebraic remainder obtained from Eq.~\eqref{appB:hessian-residual-decomposition}, not as a new independent geometric tensor.

We now restrict Eq.~\eqref{appB:hessian-residual-decomposition} to the stationary zero-current profile
\begin{equation}
p(x;\theta)=Z(\theta)^{-1}e^{-V(x;\theta)} .
\label{appB:stationary-profile}
\end{equation}
For this profile, the score is fixed by the potential and the current velocity vanishes. Using
\begin{equation}
\widetilde V
=
\frac{1}{2}B^{ij}(\partial_iV)(\partial_jV)
+
\partial_iA^i,
\qquad
A^i=-B^{ij}\partial_jV ,
\label{appB:effective-potential-drift}
\end{equation}
the potential-constraint block in Eq.~\eqref{appB:H-definition} cancels pointwise:
\begin{equation}
\mathcal{H}[p]=0 .
\label{appB:H-zero}
\end{equation}

The higher-order remnant is cancelled through its weighted form. On the stationary profile, substituting Eq.~\eqref{appB:stationary-profile} into Eq.~\eqref{appB:R-definition} and using the definition of \(\widetilde V\) together with \(A^i=-B^{ij}\partial_jV\), the surviving terms combine directly into the total-divergence form
\begin{equation}
p\mathcal{R}_{ab}
=
-\partial_i\partial_a
\left[
p\,
\partial_b(B^{ij}\partial_jV)
\right].
\label{appB:R-divergence}
\end{equation}
Thus \(\mathcal{R}_{ab}\) need not vanish pointwise. The relevant object in Eq.~\eqref{appB:hessian-residual-decomposition} is the weighted remnant \(p\mathcal{R}_{ab}\), and Eq.~\eqref{appB:R-divergence} shows that this weighted remnant is a boundary term.

Under the same boundary conditions used in the integrations by parts, Eq.~\eqref{appB:R-divergence} implies
\begin{equation}
\int_M d^dx\,p\mathcal{R}_{ab}=0 .
\label{appB:R-integrated-zero}
\end{equation}
Together with the pointwise cancellation \(\mathcal{H}[p]=0\) in Eq.~\eqref{appB:H-zero}, this removes the full residual sector in Eq.~\eqref{appB:hessian-residual-decomposition} on the stationary zero-current profile. The cancellation has two distinct parts: the \(\mathcal{H}\)-block vanishes pointwise, whereas the \(\mathcal{R}_{ab}\)-block contributes only through a weighted total divergence. The correct statement is therefore the vanishing of the integrated residual block, not the pointwise condition \(\mathcal{R}_{ab}=0\).

The same cancellation mechanism has a direct functional counterpart in the FRG discussion of Sec.~{\ref{sec:frg-entropy-lift}}. Under the correspondence summarized in {Table~\ref{tab:fp-frg-correspondence} and} Appendix~\ref{app:fp-frg-dictionary}, the probability density \(p\), the state-space derivative \(\partial_i\), and the diffusion tensor \(B^{ij}\) are lifted to the field-distribution functional \(P\), the functional derivative \(\delta/\delta\phi(x)\), and the Polchinski kernel \(C_\Lambda^{\rm Pol.}(x,y)\), respectively. Hence the residual sector in Eq.~\eqref{FRG_Constraint_Eq} is the functional analogue of the finite-dimensional residual sector derived above. In the zero-current sector at fixed RG scale, the functional potential-constraint block vanishes, while the weighted functional remnant becomes a functional total divergence. Under the same functional boundary conditions used in the integrations by parts, the integrated residual block therefore gives no contribution, leaving the curvature--Hessian balance used in the FRG formulation.

It remains to identify the Wasserstein structure behind this compatible sector.
In the JKO--Otto formulation, tangent vectors to probability space are represented by transport potentials, and the corresponding currents are gradient transport currents with respect to the Wasserstein mobility structure~\cite{Jordan1998,Otto2001,Villani2009}.
Accordingly, at fixed RG scale, the coupling-direction variation of the probability functional is represented by a functional transport potential \(\Psi_a[\phi]\) through
\begin{equation}
\frac{\partial P[\ln\Lambda,\phi;\lambda]}{\partial\lambda^a}
=
-\int_M d^d x\,
\frac{\delta}{\delta\phi(x)}
\left[
P[\ln\Lambda,\phi;\lambda]
\int_M d^d y\,
C_\Lambda^{\rm Pol.}(x,y)
\frac{\delta\Psi_a[\phi]}{\delta\phi(y)}
\right].
\label{appB:functional-transport-potential}
\end{equation}
The same definition applies to the \(b\)-direction with \(\Psi_b[\phi]\). Because the Polchinski kernel is independent of the field coordinate, it is a fixed mobility kernel on each RG slice. Under this constant-mobility convention, the pulled-back functional Wasserstein Hessian of \(\mathcal E[P]\) is
\begin{align}
\operatorname{Hess}_{\mathcal W}(\mathcal E)(\Psi_a,\Psi_b)
={}&
\int D\phi\,P\Bigg[
\int_M d^d x\,d^d y\,d^d z\,d^d w\,
C_\Lambda^{\rm Pol.}(x,z)C_\Lambda^{\rm Pol.}(y,w)
\frac{\delta^2\Psi_a}{\delta\phi(x)\delta\phi(y)}
\frac{\delta^2\Psi_b}{\delta\phi(z)\delta\phi(w)}
\nonumber\\
&\quad+
\int_M d^d x\,d^d y\,d^d z\,d^d w\,
\frac{\delta^2 V_{\rm Pol.}[\ln\Lambda,\phi]}
{\delta\phi(x)\delta\phi(y)}
C_\Lambda^{\rm Pol.}(x,z)C_\Lambda^{\rm Pol.}(y,w)
\frac{\delta\Psi_a}{\delta\phi(z)}
\frac{\delta\Psi_b}{\delta\phi(w)}
\Bigg].
\label{appB:functional-wasserstein-hessian}
\end{align}
Equation~\eqref{appB:functional-wasserstein-hessian} is the functional lift of the fixed-mobility Wasserstein Hessian. Here the word ``fixed'' refers to the mobility structure on field-configuration space at a given RG scale. It does not require \(C_\Lambda^{\rm Pol.}(x,y)\) to be local or independent of the continuous labels \(x,y\); the kernel may remain nonlocal and scale dependent. What is held fixed in the Hessian calculation is its dependence on the field configuration \(\phi\). The first term in Eq.~\eqref{appB:functional-wasserstein-hessian} is the kernel-weighted Hessian-square contribution from the entropy part of \(\mathcal E[P]\), while the second term is the Hessian of the Polchinski potential evaluated on the \(C_\Lambda^{\rm Pol.}\)-weighted transport directions. This is why the potential-Hessian term contains two Polchinski kernels.

This functional Wasserstein Hessian describes the gradient-transport sector of probability-space geometry. A general off-shell probability current may contain non-gradient, or rotational, components, but such components are not included in the Wasserstein gradient representative used in the compatible sector. The residual block in Eq.~\eqref{appB:hessian-residual-decomposition} therefore measures the obstruction to the Fisher--Wasserstein Hessian matching before this compatible restriction is imposed. The zero-current reduction derived above removes this obstruction at the integrated level, yielding the curvature--Hessian balance used in the FRG formulation.

\section{Fokker--Planck--FRG correspondence dictionary}
\label{app:fp-frg-dictionary}

{The Polchinski evolution in Section~\ref{sec:frg-fp-lift} changes the field distribution and hence the statistical distinguishability of neighboring theories. This appendix summarizes the operations that relate the construction of Section~\ref{FPE_RF} to the field-theoretic formulation. Table~\ref{tab:fp-frg-correspondence} identifies the variables, and the table below summarizes the associated operations.}

\begin{center}
\begingroup
\small
\renewcommand{\arraystretch}{1.35}
\begin{tabular}{@{}lcl@{}}
\hline\hline
\textbf{Fokker--Planck operation}
&&
\textbf{FRG operation} \\
\hline
State-space integral \(\displaystyle \int d^dx\)
& \(\longleftrightarrow\) &
Functional integral \(\displaystyle \int D\phi(x)\) \\
Index summation \(\displaystyle \sum_i\)
& \(\longleftrightarrow\) &
Continuous-label integration \(\displaystyle \int_M d^dx\) \\
State-space derivative \(\partial_i\)
& \(\longleftrightarrow\) &
Functional derivative \(\displaystyle \frac{\delta}{\delta\phi(x)}\) \\
Second derivative \(\partial_i\partial_j\)
& \(\longleftrightarrow\) &
\(\displaystyle \frac{\delta^2}{\delta\phi(x)\delta\phi(y)}\) \\
Parameter derivative \(\displaystyle \frac{\partial}{\partial\theta^a}\)
& \(\longleftrightarrow\) &
Coupling derivative \(\displaystyle \frac{\partial}{\partial\lambda^a}\) \\
Time derivative \(\partial_t\)
& \(\longleftrightarrow\) &
Scale derivative \(\displaystyle \revblue{-}\frac{d}{d\ln\Lambda}\) \\
\hline\hline
\end{tabular}
\endgroup
\end{center}

{The flow is oriented toward decreasing cutoff, so that \(t\longleftrightarrow\ln(\Lambda_0/\Lambda)\) and \(\partial_t\longleftrightarrow-d/d\ln\Lambda\). With the field configuration as the state, state-space derivatives become functional derivatives, while \(\partial_a=\partial/\partial\lambda^a\) acts on the coupling coordinates. The Polchinski kernel in Eq.~\eqref{Cuttoff_Potential} replaces the diffusion tensor in Eq.~\eqref{FP_Eq}; it is independent of the field configuration at each scale, but may be nonlocal in \(x,y\). We work in a regularized representation with a fixed integration measure, assume that coupling derivatives commute with functional integration and field derivatives, and require the probability current and its relevant derivatives to decay sufficiently rapidly to eliminate functional boundary terms.}

\section{{Details of the scalar-field example}}
\label{app:phi4-example-details}

%\begingroup
%\color{blue}

This appendix derives the initial metric deformation~\eqref{eq:phi4-initial-metric-flow} and the curvature formulas~\eqref{eq:phi4-hessian-curvature} and \eqref{eq:phi4-gaussian-curvature}, and evaluates the residual \(\mathcal I_{ab}\) in Eq.~\eqref{eq:phi4-required-hessian} for the initial two-coupling family in Section~\ref{sec:frg-phi4-example}.

\subsection{Initial RG deformation of the Fisher metric}
\label{appD:initial-metric-deformation}

We evaluate Eq.~\eqref{FRG_Metric_Dynamics} at the initial RG scale, writing \(C=C_{\Lambda_0}^{\rm Pol.}\). At this fixed scale, \(\partial_a C=\partial_a V_{\rm Pol.}=0\) for \(a\in\{r,u\}\), and \(C(x,y)=C(y,x)\). The initial family~\eqref{eq:phi4-initial-action} satisfies \(\partial_a\ln Z=-\langle O_a\rangle_0\), so
\begin{equation}
\partial_a\ln P_0=-O_a+\langle O_a\rangle_0,
\qquad
\frac{\delta}{\delta\phi(x)}\left(\partial_a\ln P_0\right)
=-\frac{\delta O_a}{\delta\phi(x)}.
\label{appD:score-field-derivatives}
\end{equation}
The expectation \(\langle O_a\rangle_0\) depends on the couplings but not on the individual field configuration being varied, so its field derivative vanishes. For the component \(\Psi(x)\) of the probability velocity~\eqref{Conserved_Current_FRG}, coupling independence of \(C\) and \(V_{\rm Pol.}\) gives
\begin{equation}
\begin{aligned}
\partial_a\Psi(x)
&=-\int_M d^dy\,C(x,y)
\frac{\delta}{\delta\phi(y)}\left(\partial_a\ln P_0\right)\\
&=\int_M d^dy\,C(x,y)\frac{\delta O_a}{\delta\phi(y)}.
\end{aligned}
\label{appD:velocity-coupling-derivative}
\end{equation}
Inserting Eqs.~\eqref{appD:score-field-derivatives} and \eqref{appD:velocity-coupling-derivative} into Eq.~\eqref{FRG_Metric_Dynamics} yields
\begin{equation}
\begin{aligned}
\dot G_{ab}
={}&-\int_M d^dx\,d^dy\,C(x,y)
\left\langle
\frac{\delta O_a}{\delta\phi(y)}\frac{\delta O_b}{\delta\phi(x)}
+\frac{\delta O_b}{\delta\phi(y)}\frac{\delta O_a}{\delta\phi(x)}
\right\rangle_0\\
={}&-2\int_M d^dx\,d^dy\,C(x,y)
\left\langle
\frac{\delta O_a}{\delta\phi(x)}\frac{\delta O_b}{\delta\phi(y)}
\right\rangle_0.
\end{aligned}
\label{appD:initial-metric-flow-derivation}
\end{equation}
The second equality follows by exchanging \(x,y\) in the first term and using the symmetry of \(C\). This establishes the first line of Eq.~\eqref{eq:phi4-initial-metric-flow} for the initial interacting distribution.

For a fixed infinitesimal coupling separation \(d\lambda=(dr,du)\), define
\begin{equation}
f_\phi(x)\equiv d\lambda^a\frac{\delta O_a}{\delta\phi(x)}
=dr\,\phi(x)+\frac{du}{6}\phi(x)^3.
\label{appD:directional-field-variation}
\end{equation}
If \(C\) is symmetric with nonnegative eigenvalues, contracting Eq.~\eqref{appD:initial-metric-flow-derivation} with \(d\lambda^a d\lambda^b\) gives
\begin{equation}
\dot G_{ab}\,d\lambda^a d\lambda^b
=-2\left\langle
\int_M d^dx\,d^dy\,f_\phi(x)C(x,y)f_\phi(y)
\right\rangle_0\le0.
\label{appD:distance-contraction}
\end{equation}
The quadratic form inside the expectation is nonnegative for every field configuration, and averaging with \(P_0\) preserves this sign. At fixed \(d\lambda\), the left-hand side of Eq.~\eqref{appD:distance-contraction} equals \(\left.\partial_\tau ds^2\right|_{\tau=0}\), where \(ds^2=G_{ab}d\lambda^a d\lambda^b\). Thus the squared statistical distance cannot increase in the first RG step. This conclusion holds for the initial interacting distribution and does not rely on the Gaussian approximation.

For \(a=b=r\), \(\delta O_r/\delta\phi(x)=\phi(x)\), so the remaining input to Eq.~\eqref{appD:initial-metric-flow-derivation} is the two-point function. Expanding its normalized expectation at fixed \(r\) about the Gaussian measure at \(u=0\) gives
\begin{equation}
\begin{aligned}
\langle\phi(x)\phi(y)\rangle_0
&=\frac{\langle\phi(x)\phi(y)e^{-uO_u}\rangle_{\rm G}}
{\langle e^{-uO_u}\rangle_{\rm G}}\\
&=D(x-y)-u\Big[
\langle\phi(x)\phi(y)O_u\rangle_{\rm G}
-D(x-y)\langle O_u\rangle_{\rm G}
\Big]+O(u^2).
\end{aligned}
\label{appD:normalized-two-point-expansion}
\end{equation}
The subtraction in Eq.~\eqref{appD:normalized-two-point-expansion} cancels the Wick contractions in which \(\phi(x)\) and \(\phi(y)\) pair with each other. For \(O_u=(4!)^{-1}\int_M d^dz\,\phi(z)^4\), the remaining contractions give \(4\times3=12\) choices for pairing the external fields with fields at \(z\); the two remaining fields at \(z\) contribute \(D(0)=I_1\). Thus \(12/4!=1/2\), and
\begin{equation}
\begin{aligned}
\langle\phi(x)\phi(y)\rangle_0
&=D(x-y)-\frac{u}{2}I_1(D^2)(x,y)+O(u^2),\\
(D^2)(x,y)
&\equiv\int_M d^dz\,D(x-z)D(z-y).
\end{aligned}
\label{appD:two-point-first-order}
\end{equation}
The convolution in Eq.~\eqref{appD:two-point-first-order} is the operator product \(D^2\). Substituting this result into Eq.~\eqref{appD:initial-metric-flow-derivation}, with traces over the regulated field modes, gives
\begin{equation}
\dot G_{rr}
=-2\operatorname{Tr}(CD)
+uI_1\operatorname{Tr}(CD^2)+O(u^2),
\label{appD:initial-rr-flow}
\end{equation}
which is the second line of Eq.~\eqref{eq:phi4-initial-metric-flow}.

\subsection{Curvature of the Hessian metric}

We write \(W(r,u)=\ln Z(r,u)\), \(W_{ab}=\partial_a\partial_b W\), and similarly for higher derivatives. All indices take values in \(\{r,u\}\), and repeated indices are summed. Coupling derivatives are taken before the Gaussian limit \(u\to0^+\).

The Fisher metric in Eq.~\eqref{eq:phi4-score-metric} has the Hessian form \(G_{ab}=W_{ab}\). Its matrix, determinant, and inverse are
\begin{equation}
\begin{gathered}
G=\begin{pmatrix}
W_{rr}&W_{ru}\\
W_{ru}&W_{uu}
\end{pmatrix},
\qquad
\det G=W_{rr}W_{uu}-W_{ru}^{\,2}>0,\\[2pt]
G^{-1}=\frac{1}{\det G}
\begin{pmatrix}
W_{uu}&-W_{ru}\\
-W_{ru}&W_{rr}
\end{pmatrix}.
\end{gathered}
\label{appD:hessian-metric}
\end{equation}
Symmetry of mixed derivatives simplifies the connection~\eqref{Connection_Torsion_Free} and gives
\begin{equation}
\begin{aligned}
\Gamma_{abc}
&=\frac12\left(\partial_aG_{bc}+\partial_bG_{ac}-\partial_cG_{ab}\right)
=\frac12W_{abc},\\
\Gamma^c_{ab}
&=\frac12G^{cd}W_{abd},
\qquad
\partial_cG^{ab}=-G^{ai}W_{ijc}G^{jb}.
\end{aligned}
\label{appD:hessian-connection}
\end{equation}
Here \(G^{ab}\) denotes an entry of the inverse matrix in Eq.~\eqref{appD:hessian-metric}. Thus the connection contains third derivatives of \(W\). The formula for \(\partial_cG^{ab}\) follows by differentiating \(G_{ai}G^{ib}=\delta_a{}^b\) with respect to \(\lambda^c\).

For the curvature sign convention in Eq.~\eqref{Ricci_Curvature_Original}, the independent Riemann component in two dimensions is obtained from Eq.~\eqref{appD:hessian-connection} as
\begin{equation}
\begin{aligned}
R_{ruru}
={}&G_{re}\left(
\partial_r\Gamma^e_{uu}-\partial_u\Gamma^e_{ru}
+\Gamma^e_{rc}\Gamma^c_{uu}-\Gamma^e_{uc}\Gamma^c_{ru}
\right)\\
={}&\underbrace{\frac12\left(W_{rruu}-W_{rruu}\right)}_{0}
+\frac14G^{cd}\left(W_{ruc}W_{rud}-W_{rrc}W_{uud}\right).
\end{aligned}
\label{appD:riemann-component}
\end{equation}
The fourth derivatives cancel because mixed derivatives commute. Derivatives of the inverse metric and products of connection coefficients leave the quadratic combination of third derivatives in Eq.~\eqref{appD:riemann-component}. The two-dimensional relation between this component and the scalar curvature then gives
\begin{equation}
R=\frac{2R_{ruru}}{\det G}
=-\frac{1}{2\det G}\,
G^{cd}\left(W_{rrc}W_{uud}-W_{ruc}W_{rud}\right).
\label{appD:scalar-contraction}
\end{equation}
Expanding the sum over \(c,d\) in Eq.~\eqref{appD:scalar-contraction} yields
\begin{equation}
\begin{aligned}
R=-\frac{1}{2\det G}\Big[
&G^{rr}\left(W_{rrr}W_{ruu}-W_{rru}^{\,2}\right)\\
&+G^{ru}\left(W_{rrr}W_{uuu}-W_{rru}W_{ruu}\right)\\
&+G^{uu}\left(W_{rru}W_{uuu}-W_{ruu}^{\,2}\right)
\Big],
\end{aligned}
\label{appD:scalar-components}
\end{equation}
where the \(c=u,d=r\) term vanishes identically. Substituting the inverse metric~\eqref{appD:hessian-metric} into Eq.~\eqref{appD:scalar-components} gives
\begin{equation}
\begin{aligned}
R=-\frac{1}{2(\det G)^2}\Big[
&W_{rr}\left(W_{rru}W_{uuu}-W_{ruu}^{\,2}\right)\\
&-W_{ru}\left(W_{rrr}W_{uuu}-W_{rru}W_{ruu}\right)\\
&+W_{uu}\left(W_{rrr}W_{ruu}-W_{rru}^{\,2}\right)
\Big].
\end{aligned}
\label{appD:scalar-expanded}
\end{equation}
The bracket in Eq.~\eqref{appD:scalar-expanded} is exactly the expansion of the \(3\times3\) determinant in Eq.~\eqref{eq:phi4-hessian-curvature} along its first row. The determinant therefore combines the three independent metric components with their first coupling derivatives. The remaining identity \(R_{ab}=(R/2)G_{ab}\) follows from the two-dimensional curvature relation \(R_{abcd}=(R/2)(G_{ac}G_{bd}-G_{ad}G_{bc})\), completing the derivation of Eq.~\eqref{eq:phi4-hessian-curvature}.

\pagebreak[0]
\begin{samepage}
\subsection{Gaussian evaluation of the curvature}

We now evaluate the coupling derivatives at \(u=0\). The identities \(\partial_rI_n=-nI_{n+1}\) and \(\partial_rJ_4=-4J_5\) follow from \(\partial_rD(p)=-D(p)^2\). Differentiating the Gaussian metric~\eqref{eq:phi4-full-gaussian-metric} with respect to \(r\) gives the first three third derivatives below. The remaining derivative satisfies \(\left.W_{uuu}\right|_{u=0}=-\langle(\delta O_u)^3\rangle_{\rm G}\). The four required derivatives are
\begin{equation}
\begin{aligned}
\left.W_{rrr}\right|_{u=0}
&=-\Omega I_3,\\
\left.W_{rru}\right|_{u=0}
&=-\Omega\left(\frac{I_2^2}{4}+\frac{I_1I_3}{2}\right),\\
\left.W_{ruu}\right|_{u=0}
&=-\Omega\left(\frac{I_1I_2^2}{4}+\frac{I_1^2I_3}{4}+\frac{J_5}{6}\right),\\
\left.W_{uuu}\right|_{u=0}
&=-\Omega\left(\frac{I_1^3I_3}{8}+\frac{3I_1^2I_2^2}{16}
+\frac{I_1J_5}{4}+\frac{J_6}{8}\right).
\end{aligned}
\label{appD:gaussian-third-derivatives}
\end{equation}
\end{samepage}
The last line follows from connected Wick contractions of three \(O_u\) insertions, including their normalization \((4!)^{-3}\). The four terms correspond to three, two, one, or no contractions within individual insertions; each such contraction gives \(I_1=D(0)\).

Substituting Eq.~\eqref{appD:gaussian-third-derivatives} and the metric~\eqref{eq:phi4-full-gaussian-metric} into the determinant in Eq.~\eqref{eq:phi4-hessian-curvature} gives
\begin{equation}
\left.
\begin{vmatrix}
W_{rr}&W_{ru}&W_{uu}\\
W_{rrr}&W_{rru}&W_{ruu}\\
W_{rru}&W_{ruu}&W_{uuu}
\end{vmatrix}
\right|_{u=0}
=\Omega^3\left[
-\frac{I_2^4J_4}{384}
+\frac{I_3J_4J_5}{144}
-\frac{I_2J_5^2}{72}
+\frac{I_2^3J_6}{64}
\right].
\label{appD:gaussian-curvature-determinant}
\end{equation}
All explicit factors of \(I_1\) cancel in Eq.~\eqref{appD:gaussian-curvature-determinant}. Dividing by \(-2(\det G)^2\), with \(\det G=\Omega^2I_2J_4/48\) from Eq.~\eqref{eq:phi4-full-gaussian-metric}, yields Eq.~\eqref{eq:phi4-gaussian-curvature}.

\subsection{Gaussian evaluation of the residual}
\label{appD:gaussian-residual}

We evaluate the residual in Eq.~\eqref{FRG_Fisher_Information_Gradient_Flow} for the initial family~\eqref{eq:phi4-initial-action}, keeping \(C=C_{\Lambda_0}^{\rm Pol.}\) and \(V_{\rm Pol.}\) independent of the couplings. All coupling derivatives are taken before the limit \(u\to0^+\). At the initial RG scale, the definition is
\begin{equation}
\begin{aligned}
\mathcal I_{ab}
&=\left.\int D\phi\,
\left[(\partial_a\partial_bP)\mathcal H_{\rm functional}
+P\mathcal R^{\rm functional}_{ab}\right]\right|_{P=P_0,\,\Lambda=\Lambda_0}\\
&=\partial_a\partial_b\mathcal F[P_0]+\frac12\dot G_{ab}.
\end{aligned}
\label{appD:residual-definition}
\end{equation}

At \(u=0\), the Gaussian propagator satisfies \(D^{-1}=(D_{\rm bare}K_{\Lambda_0})^{-1}+r\mathbf1\) in regulated-mode notation, where \(\mathbf1\) is the identity on the regulated modes. The drift potential in Eq.~\eqref{Cuttoff_Potential} contains the fixed quadratic kernel \((D_{\rm bare}K_{\Lambda_0})^{-1}=D^{-1}-r\mathbf1\), and hence \(V_{\rm Pol.}=\phi^T(D^{-1}-r\mathbf1)\phi\). Thus \(V_{\rm Pol.}\) remains independent of \(r\), although \(D\) depends on it. The translation-invariant operators \(C\) and \(D\) commute because they are diagonal in momentum space.

For \(P_{\rm G}=Z_{\rm G}^{-1}\exp[-\phi^TD^{-1}\phi/2]\), derivatives with respect to the regulated field modes are \(\partial_\phi\ln P_{\rm G}=-D^{-1}\phi\) and \(\partial_\phi V_{\rm Pol.}=2(D^{-1}-r\mathbf1)\phi\). Using \(\langle\phi_i\phi_j\rangle_{\rm G}=D_{ij}\), the three terms in Eq.~\eqref{Lifted_F_Entropy_Perfect} give, respectively,
\[
\begin{aligned}
\left.\mathcal F[P_0]\right|_{u=0}
&=\frac12\operatorname{Tr}(CD^{-1})
+2\operatorname{Tr}\!\left[C(D^{-1}-r\mathbf1)^2D\right]
-2\operatorname{Tr}\!\left[C(D^{-1}-r\mathbf1)\right]\\
&=\frac12\operatorname{Tr}(CD^{-1})
-2r\operatorname{Tr}C+2r^2\operatorname{Tr}(CD).
\end{aligned}
\]
Rearranging gives the first line below. Differentiating twice with respect to \(r\) at fixed \(C\), using \(\partial_rD^{-1}=\mathbf1\), \(\partial_rD=-D^2\), and \(\partial_r^2D=2D^3\), gives the second line:
\begin{equation}
\begin{aligned}
\left.\mathcal F[P_0]\right|_{u=0}
&=\frac12\operatorname{Tr}\!\left[C(D^{-1}-2r\mathbf1)^2D\right],\\
\left.\partial_r^2\mathcal F[P_0]\right|_{u=0}
&=4T_C-8r\operatorname{Tr}(CD^2)+4r^2\operatorname{Tr}(CD^3).
\end{aligned}
\label{appD:gaussian-entropy-hessian-rr}
\end{equation}
Combining Eq.~\eqref{appD:gaussian-entropy-hessian-rr} with \(\dot G_{rr}=-2T_C\) in Eq.~\eqref{appD:residual-definition} yields Eq.~\eqref{eq:phi4-gaussian-residual-rr}.

To obtain the mixed and quartic components, we use the functional counterparts of Eqs.~\eqref{appB:H-definition} and \eqref{appB:R-definition}. With \(C\) and \(V_{\rm Pol.}\) fixed, the weighted \(\mathcal R^{\rm functional}_{ab}\) term is a functional total divergence. The assumed boundary conditions and the score identity \(\partial_a\partial_bP_0=P_0(\delta O_a\delta O_b-G_{ab})\) reduce Eq.~\eqref{appD:residual-definition} to
\begin{equation}
\begin{aligned}
\int D\phi\,P\mathcal R^{\rm functional}_{ab}
&=-\int D\phi\int_M d^dx\,d^dy\,
\frac{\delta}{\delta\phi(x)}
\left[(\partial_bP)C(x,y)
\frac{\delta}{\delta\phi(y)}(\partial_a\ln P)\right]=0,\\
\mathcal I_{ab}
&=\left\langle(\delta O_a\delta O_b-G_{ab})
\mathcal H_{\rm functional}[P_0]\right\rangle_0.
\end{aligned}
\label{appD:residual-expectation}
\end{equation}
On this Gaussian profile, the functional version of Eq.~\eqref{appB:H-definition} becomes
\begin{equation}
\begin{aligned}
\mathcal H_{\rm functional}[P_{\rm G}]
&=\operatorname{Tr}\!\left[C(2r\mathbf1-D^{-1})\right]
+\frac12\phi^T D^{-1}\widetilde C D^{-1}\phi,\\
\widetilde C
&\equiv 3C-8rCD+4r^2CD^2.
\end{aligned}
\label{appD:residual-kernel}
\end{equation}
The field-independent trace in Eq.~\eqref{appD:residual-kernel} drops out of Eq.~\eqref{appD:residual-expectation}, since \(\langle\delta O_a\delta O_b\rangle_{\rm G}=G_{ab}\). Wick contraction of the remaining quadratic insertion with \(O_r\) and \(O_u\) gives all three components:
\begin{equation}
\begin{aligned}
\left.\mathcal I_{rr}\right|_{u=0}
&=\operatorname{Tr}(\widetilde C D),\\
\left.\mathcal I_{ru}\right|_{u=0}
=\left.\mathcal I_{ur}\right|_{u=0}
&=\frac{I_1}{2}\operatorname{Tr}(\widetilde C D)
+\frac{I_2}{4}\operatorname{Tr}\widetilde C,\\
\left.\mathcal I_{uu}\right|_{u=0}
&=\frac{I_1^2}{4}\operatorname{Tr}(\widetilde C D)
+\frac{I_1I_2}{4}\operatorname{Tr}\widetilde C\\
&\quad+\frac16\int_M d^dx\,d^dy\,
\widetilde C(x,y)D(x-y)^3.
\end{aligned}
\label{appD:gaussian-residual-components}
\end{equation}
Equation~\eqref{appD:gaussian-residual-components} expresses the residual entirely in terms of the same initial propagator and Polchinski kernel that determine the metric deformation. Its first line reproduces Eq.~\eqref{eq:phi4-gaussian-residual-rr}, while the other two retain the quartic coupling direction at the Gaussian boundary.

%\endgroup

\end{document}